\documentclass[traditabstract,longauth]{aa}  
\usepackage{txfonts}
\usepackage{bm}
\usepackage{setspace}
\usepackage{graphicx}
\usepackage[usenames]{color}
\usepackage{natbib} 
\bibpunct{(}{)}{;}{a}{}{,}

\newcommand\gracedb{GraCEDb}
\newcommand\lvalert{LVAlert}
\begin{document}

\title{First Low-Latency LIGO+Virgo Search for Binary Inspirals and their
  Electromagnetic Counterparts}

\author{
\scriptsize{
\begin{spacing}{0.01}
J.~Abadie\inst{\ref{inst1}}
\and B.~P.~Abbott\inst{\ref{inst1}}
\and R.~Abbott\inst{\ref{inst1}}
\and T.~D.~Abbott\inst{\ref{inst2}}
\and M.~Abernathy\inst{\ref{inst3}}
\and T.~Accadia\inst{\ref{inst4}}
\and F.~Acernese\inst{\ref{inst5}ac}
\and C.~Adams\inst{\ref{inst6}}
\and R.~Adhikari\inst{\ref{inst1}}
\and C.~Affeldt\inst{\ref{inst7},\ref{inst8}}
\and M.~Agathos\inst{\ref{inst9}a}
\and K.~Agatsuma\inst{\ref{inst10}}
\and P.~Ajith\inst{\ref{inst1}}
\and B.~Allen\inst{\ref{inst7},\ref{inst11},\ref{inst8}}
\and E.~Amador~Ceron\inst{\ref{inst11}}
\and D.~Amariutei\inst{\ref{inst12}}
\and S.~B.~Anderson\inst{\ref{inst1}}
\and W.~G.~Anderson\inst{\ref{inst11}}
\and K.~Arai\inst{\ref{inst1}}
\and M.~A.~Arain\inst{\ref{inst12}}
\and M.~C.~Araya\inst{\ref{inst1}}
\and S.~M.~Aston\inst{\ref{inst13}}
\and P.~Astone\inst{\ref{inst14}a}
\and D.~Atkinson\inst{\ref{inst15}}
\and P.~Aufmuth\inst{\ref{inst8},\ref{inst7}}
\and C.~Aulbert\inst{\ref{inst7},\ref{inst8}}
\and B.~E.~Aylott\inst{\ref{inst13}}
\and S.~Babak\inst{\ref{inst16}}
\and P.~Baker\inst{\ref{inst17}}
\and G.~Ballardin\inst{\ref{inst18}}
\and S.~Ballmer\inst{\ref{inst19}}
\and J.~C.~B.~Baragoya\inst{\ref{inst1}}
\and D.~Barker\inst{\ref{inst15}}
\and F.~Barone\inst{\ref{inst5}ac}
\and B.~Barr\inst{\ref{inst3}}
\and L.~Barsotti\inst{\ref{inst20}}
\and M.~Barsuglia\inst{\ref{inst21}}
\and M.~A.~Barton\inst{\ref{inst15}}
\and I.~Bartos\inst{\ref{inst22}}
\and R.~Bassiri\inst{\ref{inst3}}
\and M.~Bastarrika\inst{\ref{inst3}}
\and A.~Basti\inst{\ref{inst23}ab}
\and J.~Batch\inst{\ref{inst15}}
\and J.~Bauchrowitz\inst{\ref{inst7},\ref{inst8}}
\and Th.~S.~Bauer\inst{\ref{inst9}a}
\and M.~Bebronne\inst{\ref{inst4}}
\and D.~Beck\inst{\ref{inst24}}
\and B.~Behnke\inst{\ref{inst16}}
\and M.~Bejger\inst{\ref{inst25}c}
\and M.G.~Beker\inst{\ref{inst9}a}
\and A.~S.~Bell\inst{\ref{inst3}}
\and A.~Belletoile\inst{\ref{inst4}}
\and I.~Belopolski\inst{\ref{inst22}}
\and M.~Benacquista\inst{\ref{inst26}}
\and J.~M.~Berliner\inst{\ref{inst15}}
\and A.~Bertolini\inst{\ref{inst7},\ref{inst8}}
\and J.~Betzwieser\inst{\ref{inst1}}
\and N.~Beveridge\inst{\ref{inst3}}
\and P.~T.~Beyersdorf\inst{\ref{inst27}}
\and I.~A.~Bilenko\inst{\ref{inst28}}
\and G.~Billingsley\inst{\ref{inst1}}
\and J.~Birch\inst{\ref{inst6}}
\and R.~Biswas\inst{\ref{inst26}}
\and M.~Bitossi\inst{\ref{inst23}a}
\and M.~A.~Bizouard\inst{\ref{inst29}a}
\and E.~Black\inst{\ref{inst1}}
\and J.~K.~Blackburn\inst{\ref{inst1}}
\and L.~Blackburn\inst{\ref{inst30}}
\and D.~Blair\inst{\ref{inst31}}
\and B.~Bland\inst{\ref{inst15}}
\and M.~Blom\inst{\ref{inst9}a}
\and O.~Bock\inst{\ref{inst7},\ref{inst8}}
\and T.~P.~Bodiya\inst{\ref{inst20}}
\and C.~Bogan\inst{\ref{inst7},\ref{inst8}}
\and R.~Bondarescu\inst{\ref{inst32}}
\and F.~Bondu\inst{\ref{inst33}b}
\and L.~Bonelli\inst{\ref{inst23}ab}
\and R.~Bonnand\inst{\ref{inst34}}
\and R.~Bork\inst{\ref{inst1}}
\and M.~Born\inst{\ref{inst7},\ref{inst8}}
\and V.~Boschi\inst{\ref{inst23}a}
\and S.~Bose\inst{\ref{inst35}}
\and L.~Bosi\inst{\ref{inst36}a}
\and B. ~Bouhou\inst{\ref{inst21}}
\and S.~Braccini\inst{\ref{inst23}a}
\and C.~Bradaschia\inst{\ref{inst23}a}
\and P.~R.~Brady\inst{\ref{inst11}}
\and V.~B.~Braginsky\inst{\ref{inst28}}
\and M.~Branchesi\inst{\ref{inst37}ab}
\and J.~E.~Brau\inst{\ref{inst38}}
\and J.~Breyer\inst{\ref{inst7},\ref{inst8}}
\and T.~Briant\inst{\ref{inst39}}
\and D.~O.~Bridges\inst{\ref{inst6}}
\and A.~Brillet\inst{\ref{inst33}a}
\and M.~Brinkmann\inst{\ref{inst7},\ref{inst8}}
\and V.~Brisson\inst{\ref{inst29}a}
\and M.~Britzger\inst{\ref{inst7},\ref{inst8}}
\and A.~F.~Brooks\inst{\ref{inst1}}
\and D.~A.~Brown\inst{\ref{inst19}}
\and T.~Bulik\inst{\ref{inst25}b}
\and H.~J.~Bulten\inst{\ref{inst9}ab}
\and A.~Buonanno\inst{\ref{inst40}}
\and J.~Burguet--Castell\inst{\ref{inst11}}
\and D.~Buskulic\inst{\ref{inst4}}
\and C.~Buy\inst{\ref{inst21}}
\and R.~L.~Byer\inst{\ref{inst24}}
\and L.~Cadonati\inst{\ref{inst41}}
\and G.~Cagnoli\inst{\ref{inst37}a}
\and E.~Calloni\inst{\ref{inst5}ab}
\and J.~B.~Camp\inst{\ref{inst30}}
\and P.~Campsie\inst{\ref{inst3}}
\and J.~Cannizzo\inst{\ref{inst30}}
\and K.~Cannon\inst{\ref{inst42}}
\and B.~Canuel\inst{\ref{inst18}}
\and J.~Cao\inst{\ref{inst43}}
\and C.~D.~Capano\inst{\ref{inst19}}
\and F.~Carbognani\inst{\ref{inst18}}
\and L.~Carbone\inst{\ref{inst13}}
\and S.~Caride\inst{\ref{inst44}}
\and S.~Caudill\inst{\ref{inst45}}
\and M.~Cavagli\`a\inst{\ref{inst46}}
\and F.~Cavalier\inst{\ref{inst29}a}
\and R.~Cavalieri\inst{\ref{inst18}}
\and G.~Cella\inst{\ref{inst23}a}
\and C.~Cepeda\inst{\ref{inst1}}
\and E.~Cesarini\inst{\ref{inst37}b}
\and O.~Chaibi\inst{\ref{inst33}a}
\and T.~Chalermsongsak\inst{\ref{inst1}}
\and P.~Charlton\inst{\ref{inst47}}
\and E.~Chassande-Mottin\inst{\ref{inst21}}
\and S.~Chelkowski\inst{\ref{inst13}}
\and W.~Chen\inst{\ref{inst43}}
\and X.~Chen\inst{\ref{inst31}}
\and Y.~Chen\inst{\ref{inst48}}
\and A.~Chincarini\inst{\ref{inst49}}
\and A.~Chiummo\inst{\ref{inst18}}
\and H.~Cho\inst{\ref{inst50}}
\and J.~Chow\inst{\ref{inst51}}
\and N.~Christensen\inst{\ref{inst52}}
\and S.~S.~Y.~Chua\inst{\ref{inst51}}
\and C.~T.~Y.~Chung\inst{\ref{inst53}}
\and S.~Chung\inst{\ref{inst31}}
\and G.~Ciani\inst{\ref{inst12}}
\and D.~E.~Clark\inst{\ref{inst24}}
\and J.~Clark\inst{\ref{inst54}}
\and J.~H.~Clayton\inst{\ref{inst11}}
\and F.~Cleva\inst{\ref{inst33}a}
\and E.~Coccia\inst{\ref{inst55}ab}
\and P.-F.~Cohadon\inst{\ref{inst39}}
\and C.~N.~Colacino\inst{\ref{inst23}ab}
\and J.~Colas\inst{\ref{inst18}}
\and A.~Colla\inst{\ref{inst14}ab}
\and M.~Colombini\inst{\ref{inst14}b}
\and A.~Conte\inst{\ref{inst14}ab}
\and R.~Conte\inst{\ref{inst56}}
\and D.~Cook\inst{\ref{inst15}}
\and T.~R.~Corbitt\inst{\ref{inst20}}
\and M.~Cordier\inst{\ref{inst27}}
\and N.~Cornish\inst{\ref{inst17}}
\and A.~Corsi\inst{\ref{inst1}}
\and C.~A.~Costa\inst{\ref{inst45}}
\and M.~Coughlin\inst{\ref{inst52}}
\and J.-P.~Coulon\inst{\ref{inst33}a}
\and P.~Couvares\inst{\ref{inst19}}
\and D.~M.~Coward\inst{\ref{inst31}}
\and M.~Cowart\inst{\ref{inst6}}
\and D.~C.~Coyne\inst{\ref{inst1}}
\and J.~D.~E.~Creighton\inst{\ref{inst11}}
\and T.~D.~Creighton\inst{\ref{inst26}}
\and A.~M.~Cruise\inst{\ref{inst13}}
\and A.~Cumming\inst{\ref{inst3}}
\and L.~Cunningham\inst{\ref{inst3}}
\and E.~Cuoco\inst{\ref{inst18}}
\and R.~M.~Cutler\inst{\ref{inst13}}
\and K.~Dahl\inst{\ref{inst7},\ref{inst8}}
\and S.~L.~Danilishin\inst{\ref{inst28}}
\and R.~Dannenberg\inst{\ref{inst1}}
\and S.~D'Antonio\inst{\ref{inst55}a}
\and K.~Danzmann\inst{\ref{inst7},\ref{inst8}}
\and V.~Dattilo\inst{\ref{inst18}}
\and B.~Daudert\inst{\ref{inst1}}
\and H.~Daveloza\inst{\ref{inst26}}
\and M.~Davier\inst{\ref{inst29}a}
\and E.~J.~Daw\inst{\ref{inst57}}
\and R.~Day\inst{\ref{inst18}}
\and T.~Dayanga\inst{\ref{inst35}}
\and R.~De~Rosa\inst{\ref{inst5}ab}
\and D.~DeBra\inst{\ref{inst24}}
\and G.~Debreczeni\inst{\ref{inst58}}
\and W.~Del~Pozzo\inst{\ref{inst9}a}
\and M.~del~Prete\inst{\ref{inst59}b}
\and T.~Dent\inst{\ref{inst54}}
\and V.~Dergachev\inst{\ref{inst1}}
\and R.~DeRosa\inst{\ref{inst45}}
\and R.~DeSalvo\inst{\ref{inst1}}
\and S.~Dhurandhar\inst{\ref{inst60}}
\and L.~Di~Fiore\inst{\ref{inst5}a}
\and A.~Di~Lieto\inst{\ref{inst23}ab}
\and I.~Di~Palma\inst{\ref{inst7},\ref{inst8}}
\and M.~Di~Paolo~Emilio\inst{\ref{inst55}ac}
\and A.~Di~Virgilio\inst{\ref{inst23}a}
\and M.~D\'iaz\inst{\ref{inst26}}
\and A.~Dietz\inst{\ref{inst4}}
\and F.~Donovan\inst{\ref{inst20}}
\and K.~L.~Dooley\inst{\ref{inst12}}
\and M.~Drago\inst{\ref{inst59}ab}
\and R.~W.~P.~Drever\inst{\ref{inst61}}
\and J.~C.~Driggers\inst{\ref{inst1}}
\and Z.~Du\inst{\ref{inst43}}
\and J.-C.~Dumas\inst{\ref{inst31}}
\and T.~Eberle\inst{\ref{inst7},\ref{inst8}}
\and M.~Edgar\inst{\ref{inst3}}
\and M.~Edwards\inst{\ref{inst54}}
\and A.~Effler\inst{\ref{inst45}}
\and P.~Ehrens\inst{\ref{inst1}}
\and G.~Endr\H{o}czi\inst{\ref{inst58}}
\and R.~Engel\inst{\ref{inst1}}
\and T.~Etzel\inst{\ref{inst1}}
\and K.~Evans\inst{\ref{inst3}}
\and M.~Evans\inst{\ref{inst20}}
\and T.~Evans\inst{\ref{inst6}}
\and M.~Factourovich\inst{\ref{inst22}}
\and V.~Fafone\inst{\ref{inst55}ab}
\and S.~Fairhurst\inst{\ref{inst54}}
\and Y.~Fan\inst{\ref{inst31}}
\and B.~F.~Farr\inst{\ref{inst62}}
\and D.~Fazi\inst{\ref{inst62}}
\and H.~Fehrmann\inst{\ref{inst7},\ref{inst8}}
\and D.~Feldbaum\inst{\ref{inst12}}
\and F.~Feroz\inst{\ref{inst63}}
\and I.~Ferrante\inst{\ref{inst23}ab}
\and F.~Fidecaro\inst{\ref{inst23}ab}
\and L.~S.~Finn\inst{\ref{inst32}}
\and I.~Fiori\inst{\ref{inst18}}
\and R.~P.~Fisher\inst{\ref{inst32}}
\and R.~Flaminio\inst{\ref{inst34}}
\and M.~Flanigan\inst{\ref{inst15}}
\and S.~Foley\inst{\ref{inst20}}
\and E.~Forsi\inst{\ref{inst6}}
\and L.~A.~Forte\inst{\ref{inst5}a}
\and N.~Fotopoulos\inst{\ref{inst1}}
\and J.-D.~Fournier\inst{\ref{inst33}a}
\and J.~Franc\inst{\ref{inst34}}
\and S.~Frasca\inst{\ref{inst14}ab}
\and F.~Frasconi\inst{\ref{inst23}a}
\and M.~Frede\inst{\ref{inst7},\ref{inst8}}
\and M.~Frei\inst{\ref{inst64},\ref{inst85}}
\and Z.~Frei\inst{\ref{inst65}}
\and A.~Freise\inst{\ref{inst13}}
\and R.~Frey\inst{\ref{inst38}}
\and T.~T.~Fricke\inst{\ref{inst45}}
\and D.~Friedrich\inst{\ref{inst7},\ref{inst8}}
\and P.~Fritschel\inst{\ref{inst20}}
\and V.~V.~Frolov\inst{\ref{inst6}}
\and M.-K.~Fujimoto\inst{\ref{inst10}}
\and P.~J.~Fulda\inst{\ref{inst13}}
\and M.~Fyffe\inst{\ref{inst6}}
\and J.~Gair\inst{\ref{inst63}}
\and M.~Galimberti\inst{\ref{inst34}}
\and L.~Gammaitoni\inst{\ref{inst36}ab}
\and J.~Garcia\inst{\ref{inst15}}
\and F.~Garufi\inst{\ref{inst5}ab}
\and M.~E.~G\'asp\'ar\inst{\ref{inst58}}
\and G.~Gemme\inst{\ref{inst49}}
\and R.~Geng\inst{\ref{inst43}}
\and E.~Genin\inst{\ref{inst18}}
\and A.~Gennai\inst{\ref{inst23}a}
\and L.~\'A.~Gergely\inst{\ref{inst66}}
\and S.~Ghosh\inst{\ref{inst35}}
\and J.~A.~Giaime\inst{\ref{inst45},\ref{inst6}}
\and S.~Giampanis\inst{\ref{inst11}}
\and K.~D.~Giardina\inst{\ref{inst6}}
\and A.~Giazotto\inst{\ref{inst23}a}
\and S.~Gil\inst{\ref{inst67}}
\and C.~Gill\inst{\ref{inst3}}
\and J.~Gleason\inst{\ref{inst12}}
\and E.~Goetz\inst{\ref{inst7},\ref{inst8}}
\and L.~M.~Goggin\inst{\ref{inst11}}
\and G.~Gonz\'alez\inst{\ref{inst45}}
\and M.~L.~Gorodetsky\inst{\ref{inst28}}
\and S.~Go{\ss}ler\inst{\ref{inst7},\ref{inst8}}
\and R.~Gouaty\inst{\ref{inst4}}
\and C.~Graef\inst{\ref{inst7},\ref{inst8}}
\and P.~B.~Graff\inst{\ref{inst63}}
\and M.~Granata\inst{\ref{inst21}}
\and A.~Grant\inst{\ref{inst3}}
\and S.~Gras\inst{\ref{inst31}}
\and C.~Gray\inst{\ref{inst15}}
\and N.~Gray\inst{\ref{inst3}}
\and R.~J.~S.~Greenhalgh\inst{\ref{inst68}}
\and A.~M.~Gretarsson\inst{\ref{inst69}}
\and C.~Greverie\inst{\ref{inst33}a}
\and R.~Grosso\inst{\ref{inst26}}
\and H.~Grote\inst{\ref{inst7},\ref{inst8}}
\and S.~Grunewald\inst{\ref{inst16}}
\and G.~M.~Guidi\inst{\ref{inst37}ab}
\and R.~Gupta\inst{\ref{inst60}}
\and E.~K.~Gustafson\inst{\ref{inst1}}
\and R.~Gustafson\inst{\ref{inst44}}
\and T.~Ha\inst{\ref{inst70}}
\and J.~M.~Hallam\inst{\ref{inst13}}
\and D.~Hammer\inst{\ref{inst11}}
\and G.~Hammond\inst{\ref{inst3}}
\and J.~Hanks\inst{\ref{inst15}}
\and C.~Hanna\inst{\ref{inst1},\ref{inst71}}
\and J.~Hanson\inst{\ref{inst6}}
\and J.~Harms\inst{\ref{inst61}}
\and G.~M.~Harry\inst{\ref{inst20}}
\and I.~W.~Harry\inst{\ref{inst54}}
\and E.~D.~Harstad\inst{\ref{inst38}}
\and M.~T.~Hartman\inst{\ref{inst12}}
\and K.~Haughian\inst{\ref{inst3}}
\and K.~Hayama\inst{\ref{inst10}}
\and J.-F.~Hayau\inst{\ref{inst33}b}
\and J.~Heefner\inst{\ref{inst1}}
\and A.~Heidmann\inst{\ref{inst39}}
\and M.~C.~Heintze\inst{\ref{inst12}}
\and H.~Heitmann\inst{\ref{inst33}}
\and P.~Hello\inst{\ref{inst29}a}
\and M.~A.~Hendry\inst{\ref{inst3}}
\and I.~S.~Heng\inst{\ref{inst3}}
\and A.~W.~Heptonstall\inst{\ref{inst1}}
\and V.~Herrera\inst{\ref{inst24}}
\and M.~Hewitson\inst{\ref{inst7},\ref{inst8}}
\and S.~Hild\inst{\ref{inst3}}
\and D.~Hoak\inst{\ref{inst41}}
\and K.~A.~Hodge\inst{\ref{inst1}}
\and K.~Holt\inst{\ref{inst6}}
\and M.~Holtrop\inst{\ref{inst72}}
\and T.~Hong\inst{\ref{inst48}}
\and S.~Hooper\inst{\ref{inst31}}
\and D.~J.~Hosken\inst{\ref{inst73}}
\and J.~Hough\inst{\ref{inst3}}
\and E.~J.~Howell\inst{\ref{inst31}}
\and B.~Hughey\inst{\ref{inst11}}
\and S.~Husa\inst{\ref{inst67}}
\and S.~H.~Huttner\inst{\ref{inst3}}
\and R.~Inta\inst{\ref{inst51}}
\and T.~Isogai\inst{\ref{inst52}}
\and A.~Ivanov\inst{\ref{inst1}}
\and K.~Izumi\inst{\ref{inst10}}
\and M.~Jacobson\inst{\ref{inst1}}
\and E.~James\inst{\ref{inst1}}
\and Y.~J.~Jang\inst{\ref{inst43}}
\and P.~Jaranowski\inst{\ref{inst25}d}
\and E.~Jesse\inst{\ref{inst69}}
\and W.~W.~Johnson\inst{\ref{inst45}}
\and D.~I.~Jones\inst{\ref{inst74}}
\and G.~Jones\inst{\ref{inst54}}
\and R.~Jones\inst{\ref{inst3}}
\and L.~Ju\inst{\ref{inst31}}
\and P.~Kalmus\inst{\ref{inst1}}
\and V.~Kalogera\inst{\ref{inst62}}
\and S.~Kandhasamy\inst{\ref{inst75}}
\and G.~Kang\inst{\ref{inst76}}
\and J.~B.~Kanner\inst{\ref{inst40}}
\and R.~Kasturi\inst{\ref{inst77}}
\and E.~Katsavounidis\inst{\ref{inst20}}
\and W.~Katzman\inst{\ref{inst6}}
\and H.~Kaufer\inst{\ref{inst7},\ref{inst8}}
\and K.~Kawabe\inst{\ref{inst15}}
\and S.~Kawamura\inst{\ref{inst10}}
\and F.~Kawazoe\inst{\ref{inst7},\ref{inst8}}
\and D.~Kelley\inst{\ref{inst19}}
\and W.~Kells\inst{\ref{inst1}}
\and D.~G.~Keppel\inst{\ref{inst1}}
\and Z.~Keresztes\inst{\ref{inst66}}
\and A.~Khalaidovski\inst{\ref{inst7},\ref{inst8}}
\and F.~Y.~Khalili\inst{\ref{inst28}}
\and E.~A.~Khazanov\inst{\ref{inst78}}
\and B.~Kim\inst{\ref{inst76}}
\and C.~Kim\inst{\ref{inst79}}
\and H.~Kim\inst{\ref{inst7},\ref{inst8}}
\and K.~Kim\inst{\ref{inst80}}
\and N.~Kim\inst{\ref{inst24}}
\and Y.~-M.~Kim\inst{\ref{inst50}}
\and P.~J.~King\inst{\ref{inst1}}
\and D.~L.~Kinzel\inst{\ref{inst6}}
\and J.~S.~Kissel\inst{\ref{inst20}}
\and S.~Klimenko\inst{\ref{inst12}}
\and K.~Kokeyama\inst{\ref{inst13}}
\and V.~Kondrashov\inst{\ref{inst1}}
\and S.~Koranda\inst{\ref{inst11}}
\and W.~Z.~Korth\inst{\ref{inst1}}
\and I.~Kowalska\inst{\ref{inst25}b}
\and D.~Kozak\inst{\ref{inst1}}
\and O.~Kranz\inst{\ref{inst7},\ref{inst8}}
\and V.~Kringel\inst{\ref{inst7},\ref{inst8}}
\and S.~Krishnamurthy\inst{\ref{inst62}}
\and B.~Krishnan\inst{\ref{inst16}}
\and A.~Kr\'olak\inst{\ref{inst25}ae}
\and G.~Kuehn\inst{\ref{inst7},\ref{inst8}}
\and R.~Kumar\inst{\ref{inst3}}
\and P.~Kwee\inst{\ref{inst8},\ref{inst7}}
\and P.~K.~Lam\inst{\ref{inst51}}
\and M.~Landry\inst{\ref{inst15}}
\and B.~Lantz\inst{\ref{inst24}}
\and N.~Lastzka\inst{\ref{inst7},\ref{inst8}}
\and C.~Lawrie\inst{\ref{inst3}}
\and A.~Lazzarini\inst{\ref{inst1}}
\and P.~Leaci\inst{\ref{inst16}}
\and C.~H.~Lee\inst{\ref{inst50}}
\and H.~K.~Lee\inst{\ref{inst80}}
\and H.~M.~Lee\inst{\ref{inst81}}
\and J.~R.~Leong\inst{\ref{inst7},\ref{inst8}}
\and I.~Leonor\inst{\ref{inst38}}
\and N.~Leroy\inst{\ref{inst29}a}
\and N.~Letendre\inst{\ref{inst4}}
\and J.~Li\inst{\ref{inst43}}
\and T.~G.~F.~Li\inst{\ref{inst9}a}
\and N.~Liguori\inst{\ref{inst59}ab}
\and P.~E.~Lindquist\inst{\ref{inst1}}
\and Y.~Liu\inst{\ref{inst43}}
\and Z.~Liu\inst{\ref{inst12}}
\and N.~A.~Lockerbie\inst{\ref{inst82}}
\and D.~Lodhia\inst{\ref{inst13}}
\and M.~Lorenzini\inst{\ref{inst37}a}
\and V.~Loriette\inst{\ref{inst29}b}
\and M.~Lormand\inst{\ref{inst6}}
\and G.~Losurdo\inst{\ref{inst37}a}
\and J.~Lough\inst{\ref{inst19}}
\and J.~Luan\inst{\ref{inst48}}
\and M.~Lubinski\inst{\ref{inst15}}
\and H.~L\"uck\inst{\ref{inst7},\ref{inst8}}
\and A.~P.~Lundgren\inst{\ref{inst32}}
\and E.~Macdonald\inst{\ref{inst3}}
\and B.~Machenschalk\inst{\ref{inst7},\ref{inst8}}
\and M.~MacInnis\inst{\ref{inst20}}
\and D.~M.~Macleod\inst{\ref{inst54}}
\and M.~Mageswaran\inst{\ref{inst1}}
\and K.~Mailand\inst{\ref{inst1}}
\and E.~Majorana\inst{\ref{inst14}a}
\and I.~Maksimovic\inst{\ref{inst29}b}
\and N.~Man\inst{\ref{inst33}a}
\and I.~Mandel\inst{\ref{inst20}}
\and V.~Mandic\inst{\ref{inst75}}
\and M.~Mantovani\inst{\ref{inst23}ac}
\and A.~Marandi\inst{\ref{inst24}}
\and F.~Marchesoni\inst{\ref{inst36}a}
\and F.~Marion\inst{\ref{inst4}}
\and S.~M\'arka\inst{\ref{inst22}}
\and Z.~M\'arka\inst{\ref{inst22}}
\and A.~Markosyan\inst{\ref{inst24}}
\and E.~Maros\inst{\ref{inst1}}
\and J.~Marque\inst{\ref{inst18}}
\and F.~Martelli\inst{\ref{inst37}ab}
\and I.~W.~Martin\inst{\ref{inst3}}
\and R.~M.~Martin\inst{\ref{inst12}}
\and J.~N.~Marx\inst{\ref{inst1}}
\and K.~Mason\inst{\ref{inst20}}
\and A.~Masserot\inst{\ref{inst4}}
\and F.~Matichard\inst{\ref{inst20}}
\and L.~Matone\inst{\ref{inst22}}
\and R.~A.~Matzner\inst{\ref{inst64}}
\and N.~Mavalvala\inst{\ref{inst20}}
\and G.~Mazzolo\inst{\ref{inst7},\ref{inst8}}
\and R.~McCarthy\inst{\ref{inst15}}
\and D.~E.~McClelland\inst{\ref{inst51}}
\and S.~C.~McGuire\inst{\ref{inst83}}
\and G.~McIntyre\inst{\ref{inst1}}
\and J.~McIver\inst{\ref{inst41}}
\and D.~J.~A.~McKechan\inst{\ref{inst54}}
\and S.~McWilliams\inst{\ref{inst22}}
\and G.~D.~Meadors\inst{\ref{inst44}}
\and M.~Mehmet\inst{\ref{inst7},\ref{inst8}}
\and T.~Meier\inst{\ref{inst8},\ref{inst7}}
\and A.~Melatos\inst{\ref{inst53}}
\and A.~C.~Melissinos\inst{\ref{inst84}}
\and G.~Mendell\inst{\ref{inst15}}
\and R.~A.~Mercer\inst{\ref{inst11}}
\and S.~Meshkov\inst{\ref{inst1}}
\and C.~Messenger\inst{\ref{inst54}}
\and M.~S.~Meyer\inst{\ref{inst6}}
\and C.~Michel\inst{\ref{inst34}}
\and L.~Milano\inst{\ref{inst5}ab}
\and J.~Miller\inst{\ref{inst51}}
\and Y.~Minenkov\inst{\ref{inst55}a}
\and V.~P.~Mitrofanov\inst{\ref{inst28}}
\and G.~Mitselmakher\inst{\ref{inst12}}
\and R.~Mittleman\inst{\ref{inst20}}
\and O.~Miyakawa\inst{\ref{inst10}}
\and B.~Moe\inst{\ref{inst11}}
\and M.~Mohan\inst{\ref{inst18}}
\and S.~D.~Mohanty\inst{\ref{inst26}}
\and S.~R.~P.~Mohapatra\inst{\ref{inst41}}
\and G.~Moreno\inst{\ref{inst15}}
\and N.~Morgado\inst{\ref{inst34}}
\and A.~Morgia\inst{\ref{inst55}ab}
\and T.~Mori\inst{\ref{inst10}}
\and S.~R.~Morriss\inst{\ref{inst26}}
\and S.~Mosca\inst{\ref{inst5}ab}
\and K.~Mossavi\inst{\ref{inst7},\ref{inst8}}
\and B.~Mours\inst{\ref{inst4}}
\and C.~M.~Mow--Lowry\inst{\ref{inst51}}
\and C.~L.~Mueller\inst{\ref{inst12}}
\and G.~Mueller\inst{\ref{inst12}}
\and S.~Mukherjee\inst{\ref{inst26}}
\and A.~Mullavey\inst{\ref{inst51}}
\and H.~M\"uller-Ebhardt\inst{\ref{inst7},\ref{inst8}}
\and J.~Munch\inst{\ref{inst73}}
\and D.~Murphy\inst{\ref{inst22}}
\and P.~G.~Murray\inst{\ref{inst3}}
\and A.~Mytidis\inst{\ref{inst12}}
\and T.~Nash\inst{\ref{inst1}}
\and L.~Naticchioni\inst{\ref{inst14}ab}
\and V.~Necula\inst{\ref{inst12}}
\and J.~Nelson\inst{\ref{inst3}}
\and G.~Newton\inst{\ref{inst3}}
\and T.~Nguyen\inst{\ref{inst51}}
\and A.~Nishizawa\inst{\ref{inst10}}
\and A.~Nitz\inst{\ref{inst19}}
\and F.~Nocera\inst{\ref{inst18}}
\and D.~Nolting\inst{\ref{inst6}}
\and M.~E.~Normandin\inst{\ref{inst26}}
\and L.~Nuttall\inst{\ref{inst54}}
\and E.~Ochsner\inst{\ref{inst40}}
\and J.~O'Dell\inst{\ref{inst68}}
\and E.~Oelker\inst{\ref{inst20}}
\and G.~H.~Ogin\inst{\ref{inst1}}
\and J.~J.~Oh\inst{\ref{inst70}}
\and S.~H.~Oh\inst{\ref{inst70}}
\and B.~O'Reilly\inst{\ref{inst6}}
\and R.~O'Shaughnessy\inst{\ref{inst11}}
\and C.~Osthelder\inst{\ref{inst1}}
\and C.~D.~Ott\inst{\ref{inst48}}
\and D.~J.~Ottaway\inst{\ref{inst73}}
\and R.~S.~Ottens\inst{\ref{inst12}}
\and H.~Overmier\inst{\ref{inst6}}
\and B.~J.~Owen\inst{\ref{inst32}}
\and A.~Page\inst{\ref{inst13}}
\and G.~Pagliaroli\inst{\ref{inst55}ac}
\and L.~Palladino\inst{\ref{inst55}ac}
\and C.~Palomba\inst{\ref{inst14}a}
\and Y.~Pan\inst{\ref{inst40}}
\and C.~Pankow\inst{\ref{inst12}}
\and F.~Paoletti\inst{\ref{inst23}a,\ref{inst18}}
\and M.~A.~Papa\inst{\ref{inst16},\ref{inst11}}
\and M.~Parisi\inst{\ref{inst5}ab}
\and A.~Pasqualetti\inst{\ref{inst18}}
\and R.~Passaquieti\inst{\ref{inst23}ab}
\and D.~Passuello\inst{\ref{inst23}a}
\and P.~Patel\inst{\ref{inst1}}
\and M.~Pedraza\inst{\ref{inst1}}
\and P.~Peiris\inst{\ref{inst85}}
\and L.~Pekowsky\inst{\ref{inst19}}
\and S.~Penn\inst{\ref{inst77}}
\and A.~Perreca\inst{\ref{inst19}}
\and G.~Persichetti\inst{\ref{inst5}ab}
\and M.~Phelps\inst{\ref{inst1}}
\and M.~Pickenpack\inst{\ref{inst7},\ref{inst8}}
\and F.~Piergiovanni\inst{\ref{inst37}ab}
\and M.~Pietka\inst{\ref{inst25}d}
\and L.~Pinard\inst{\ref{inst34}}
\and I.~M.~Pinto\inst{\ref{inst86}}
\and M.~Pitkin\inst{\ref{inst3}}
\and H.~J.~Pletsch\inst{\ref{inst7},\ref{inst8}}
\and M.~V.~Plissi\inst{\ref{inst3}}
\and R.~Poggiani\inst{\ref{inst23}ab}
\and J.~P\"old\inst{\ref{inst7},\ref{inst8}}
\and F.~Postiglione\inst{\ref{inst56}}
\and M.~Prato\inst{\ref{inst49}}
\and V.~Predoi\inst{\ref{inst54}}
\and T.~Prestegard\inst{\ref{inst75}}
\and L.~R.~Price\inst{\ref{inst1}}
\and M.~Prijatelj\inst{\ref{inst7},\ref{inst8}}
\and M.~Principe\inst{\ref{inst86}}
\and S.~Privitera\inst{\ref{inst1}}
\and R.~Prix\inst{\ref{inst7},\ref{inst8}}
\and G.~A.~Prodi\inst{\ref{inst59}ab}
\and L.~G.~Prokhorov\inst{\ref{inst28}}
\and O.~Puncken\inst{\ref{inst7},\ref{inst8}}
\and M.~Punturo\inst{\ref{inst36}a}
\and P.~Puppo\inst{\ref{inst14}a}
\and V.~Quetschke\inst{\ref{inst26}}
\and R.~Quitzow-James\inst{\ref{inst38}}
\and F.~J.~Raab\inst{\ref{inst15}}
\and D.~S.~Rabeling\inst{\ref{inst9}ab}
\and I.~R\'acz\inst{\ref{inst58}}
\and H.~Radkins\inst{\ref{inst15}}
\and P.~Raffai\inst{\ref{inst65}}
\and M.~Rakhmanov\inst{\ref{inst26}}
\and B.~Rankins\inst{\ref{inst46}}
\and P.~Rapagnani\inst{\ref{inst14}ab}
\and V.~Raymond\inst{\ref{inst62}}
\and V.~Re\inst{\ref{inst55}ab}
\and K.~Redwine\inst{\ref{inst22}}
\and C.~M.~Reed\inst{\ref{inst15}}
\and T.~Reed\inst{\ref{inst87}}
\and T.~Regimbau\inst{\ref{inst33}a}
\and S.~Reid\inst{\ref{inst3}}
\and D.~H.~Reitze\inst{\ref{inst12}}
\and F.~Ricci\inst{\ref{inst14}ab}
\and R.~Riesen\inst{\ref{inst6}}
\and K.~Riles\inst{\ref{inst44}}
\and N.~A.~Robertson\inst{\ref{inst1},\ref{inst3}}
\and F.~Robinet\inst{\ref{inst29}a}
\and C.~Robinson\inst{\ref{inst54}}
\and E.~L.~Robinson\inst{\ref{inst16}}
\and A.~Rocchi\inst{\ref{inst55}a}
\and S.~Roddy\inst{\ref{inst6}}
\and C.~Rodriguez\inst{\ref{inst62}}
\and M.~Rodruck\inst{\ref{inst15}}
\and L.~Rolland\inst{\ref{inst4}}
\and J.~G.~Rollins\inst{\ref{inst1}}
\and J.~D.~Romano\inst{\ref{inst26}}
\and R.~Romano\inst{\ref{inst5}ac}
\and J.~H.~Romie\inst{\ref{inst6}}
\and D.~Rosi\'nska\inst{\ref{inst25}cf}
\and C.~R\"{o}ver\inst{\ref{inst7},\ref{inst8}}
\and S.~Rowan\inst{\ref{inst3}}
\and A.~R\"udiger\inst{\ref{inst7},\ref{inst8}}
\and P.~Ruggi\inst{\ref{inst18}}
\and K.~Ryan\inst{\ref{inst15}}
\and P.~Sainathan\inst{\ref{inst12}}
\and F.~Salemi\inst{\ref{inst7},\ref{inst8}}
\and L.~Sammut\inst{\ref{inst53}}
\and V.~Sandberg\inst{\ref{inst15}}
\and V.~Sannibale\inst{\ref{inst1}}
\and L.~Santamar\'ia\inst{\ref{inst1}}
\and I.~Santiago-Prieto\inst{\ref{inst3}}
\and G.~Santostasi\inst{\ref{inst88}}
\and B.~Sassolas\inst{\ref{inst34}}
\and B.~S.~Sathyaprakash\inst{\ref{inst54}}
\and S.~Sato\inst{\ref{inst10}}
\and P.~R.~Saulson\inst{\ref{inst19}}
\and R.~L.~Savage\inst{\ref{inst15}}
\and R.~Schilling\inst{\ref{inst7},\ref{inst8}}
\and R.~Schnabel\inst{\ref{inst7},\ref{inst8}}
\and R.~M.~S.~Schofield\inst{\ref{inst38}}
\and E.~Schreiber\inst{\ref{inst7},\ref{inst8}}
\and B.~Schulz\inst{\ref{inst7},\ref{inst8}}
\and B.~F.~Schutz\inst{\ref{inst16},\ref{inst54}}
\and P.~Schwinberg\inst{\ref{inst15}}
\and J.~Scott\inst{\ref{inst3}}
\and S.~M.~Scott\inst{\ref{inst51}}
\and F.~Seifert\inst{\ref{inst1}}
\and D.~Sellers\inst{\ref{inst6}}
\and D.~Sentenac\inst{\ref{inst18}}
\and A.~Sergeev\inst{\ref{inst78}}
\and D.~A.~Shaddock\inst{\ref{inst51}}
\and M.~Shaltev\inst{\ref{inst7},\ref{inst8}}
\and B.~Shapiro\inst{\ref{inst20}}
\and P.~Shawhan\inst{\ref{inst40}}
\and D.~H.~Shoemaker\inst{\ref{inst20}}
\and A.~Sibley\inst{\ref{inst6}}
\and X.~Siemens\inst{\ref{inst11}}
\and D.~Sigg\inst{\ref{inst15}}
\and A.~Singer\inst{\ref{inst1}}
\and L.~Singer\inst{\ref{inst1}}
\and A.~M.~Sintes\inst{\ref{inst67}}
\and G.~R.~Skelton\inst{\ref{inst11}}
\and B.~J.~J.~Slagmolen\inst{\ref{inst51}}
\and J.~Slutsky\inst{\ref{inst45}}
\and J.~R.~Smith\inst{\ref{inst2}}
\and M.~R.~Smith\inst{\ref{inst1}}
\and R.~J.~E.~Smith\inst{\ref{inst13}}
\and N.~D.~Smith-Lefebvre\inst{\ref{inst15}}
\and K.~Somiya\inst{\ref{inst48}}
\and B.~Sorazu\inst{\ref{inst3}}
\and J.~Soto\inst{\ref{inst20}}
\and F.~C.~Speirits\inst{\ref{inst3}}
\and L.~Sperandio\inst{\ref{inst55}ab}
\and M.~Stefszky\inst{\ref{inst51}}
\and A.~J.~Stein\inst{\ref{inst20}}
\and L.~C.~Stein\inst{\ref{inst20}}
\and E.~Steinert\inst{\ref{inst15}}
\and J.~Steinlechner\inst{\ref{inst7},\ref{inst8}}
\and S.~Steinlechner\inst{\ref{inst7},\ref{inst8}}
\and S.~Steplewski\inst{\ref{inst35}}
\and A.~Stochino\inst{\ref{inst1}}
\and R.~Stone\inst{\ref{inst26}}
\and K.~A.~Strain\inst{\ref{inst3}}
\and S.~E.~Strigin\inst{\ref{inst28}}
\and A.~S.~Stroeer\inst{\ref{inst26}}
\and R.~Sturani\inst{\ref{inst37}ab}
\and A.~L.~Stuver\inst{\ref{inst6}}
\and T.~Z.~Summerscales\inst{\ref{inst89}}
\and M.~Sung\inst{\ref{inst45}}
\and S.~Susmithan\inst{\ref{inst31}}
\and P.~J.~Sutton\inst{\ref{inst54}}
\and B.~Swinkels\inst{\ref{inst18}}
\and M.~Tacca\inst{\ref{inst18}}
\and L.~Taffarello\inst{\ref{inst59}c}
\and D.~Talukder\inst{\ref{inst35}}
\and D.~B.~Tanner\inst{\ref{inst12}}
\and S.~P.~Tarabrin\inst{\ref{inst7},\ref{inst8}}
\and J.~R.~Taylor\inst{\ref{inst7},\ref{inst8}}
\and R.~Taylor\inst{\ref{inst1}}
\and P.~Thomas\inst{\ref{inst15}}
\and K.~A.~Thorne\inst{\ref{inst6}}
\and K.~S.~Thorne\inst{\ref{inst48}}
\and E.~Thrane\inst{\ref{inst75}}
\and A.~Th\"uring\inst{\ref{inst8},\ref{inst7}}
\and K.~V.~Tokmakov\inst{\ref{inst82}}
\and C.~Tomlinson\inst{\ref{inst57}}
\and A.~Toncelli\inst{\ref{inst23}ab}
\and M.~Tonelli\inst{\ref{inst23}ab}
\and O.~Torre\inst{\ref{inst23}ac}
\and C.~Torres\inst{\ref{inst6}}
\and C.~I.~Torrie\inst{\ref{inst1},\ref{inst3}}
\and E.~Tournefier\inst{\ref{inst4}}
\and F.~Travasso\inst{\ref{inst36}ab}
\and G.~Traylor\inst{\ref{inst6}}
\and K.~Tseng\inst{\ref{inst24}}
\and D.~Ugolini\inst{\ref{inst90}}
\and H.~Vahlbruch\inst{\ref{inst8},\ref{inst7}}
\and G.~Vajente\inst{\ref{inst23}ab}
\and J.~F.~J.~van~den~Brand\inst{\ref{inst9}ab}
\and C.~Van~Den~Broeck\inst{\ref{inst9}a}
\and S.~van~der~Putten\inst{\ref{inst9}a}
\and A.~A.~van~Veggel\inst{\ref{inst3}}
\and S.~Vass\inst{\ref{inst1}}
\and M.~Vasuth\inst{\ref{inst58}}
\and R.~Vaulin\inst{\ref{inst20}}
\and M.~Vavoulidis\inst{\ref{inst29}a}
\and A.~Vecchio\inst{\ref{inst13}}
\and G.~Vedovato\inst{\ref{inst59}c}
\and J.~Veitch\inst{\ref{inst54}}
\and P.~J.~Veitch\inst{\ref{inst73}}
\and C.~Veltkamp\inst{\ref{inst7},\ref{inst8}}
\and D.~Verkindt\inst{\ref{inst4}}
\and F.~Vetrano\inst{\ref{inst37}ab}
\and A.~Vicer\'e\inst{\ref{inst37}ab}
\and A.~E.~Villar\inst{\ref{inst1}}
\and J.-Y.~Vinet\inst{\ref{inst33}a}
\and S.~Vitale\inst{\ref{inst69}}
\and S.~Vitale\inst{\ref{inst9}a}
\and H.~Vocca\inst{\ref{inst36}a}
\and C.~Vorvick\inst{\ref{inst15}}
\and S.~P.~Vyatchanin\inst{\ref{inst28}}
\and A.~Wade\inst{\ref{inst51}}
\and L.~Wade\inst{\ref{inst11}}
\and M.~Wade\inst{\ref{inst11}}
\and S.~J.~Waldman\inst{\ref{inst20}}
\and L.~Wallace\inst{\ref{inst1}}
\and Y.~Wan\inst{\ref{inst43}}
\and M.~Wang\inst{\ref{inst13}}
\and X.~Wang\inst{\ref{inst43}}
\and Z.~Wang\inst{\ref{inst43}}
\and A.~Wanner\inst{\ref{inst7},\ref{inst8}}
\and R.~L.~Ward\inst{\ref{inst21}}
\and M.~Was\inst{\ref{inst29}a}
\and M.~Weinert\inst{\ref{inst7},\ref{inst8}}
\and A.~J.~Weinstein\inst{\ref{inst1}}
\and R.~Weiss\inst{\ref{inst20}}
\and L.~Wen\inst{\ref{inst48},\ref{inst31}}
\and P.~Wessels\inst{\ref{inst7},\ref{inst8}}
\and M.~West\inst{\ref{inst19}}
\and T.~Westphal\inst{\ref{inst7},\ref{inst8}}
\and K.~Wette\inst{\ref{inst7},\ref{inst8}}
\and J.~T.~Whelan\inst{\ref{inst85}}
\and S.~E.~Whitcomb\inst{\ref{inst1},\ref{inst31}}
\and D.~J.~White\inst{\ref{inst57}}
\and B.~F.~Whiting\inst{\ref{inst12}}
\and C.~Wilkinson\inst{\ref{inst15}}
\and P.~A.~Willems\inst{\ref{inst1}}
\and L.~Williams\inst{\ref{inst12}}
\and R.~Williams\inst{\ref{inst1}}
\and B.~Willke\inst{\ref{inst7},\ref{inst8}}
\and L.~Winkelmann\inst{\ref{inst7},\ref{inst8}}
\and W.~Winkler\inst{\ref{inst7},\ref{inst8}}
\and C.~C.~Wipf\inst{\ref{inst20}}
\and A.~G.~Wiseman\inst{\ref{inst11}}
\and H.~Wittel\inst{\ref{inst7},\ref{inst8}}
\and G.~Woan\inst{\ref{inst3}}
\and R.~Wooley\inst{\ref{inst6}}
\and J.~Worden\inst{\ref{inst15}}
\and I.~Yakushin\inst{\ref{inst6}}
\and H.~Yamamoto\inst{\ref{inst1}}
\and K.~Yamamoto\inst{\ref{inst7},\ref{inst8},\ref{inst59}bd}
\and C.~C.~Yancey\inst{\ref{inst40}}
\and H.~Yang\inst{\ref{inst48}}
\and D.~Yeaton-Massey\inst{\ref{inst1}}
\and S.~Yoshida\inst{\ref{inst91}}
\and P.~Yu\inst{\ref{inst11}}
\and M.~Yvert\inst{\ref{inst4}}
\and A.~Zadro\'zny\inst{\ref{inst25}e}
\and M.~Zanolin\inst{\ref{inst69}}
\and J.-P.~Zendri\inst{\ref{inst59}c}
\and F.~Zhang\inst{\ref{inst43}}
\and L.~Zhang\inst{\ref{inst1}}
\and W.~Zhang\inst{\ref{inst43}}
\and C.~Zhao\inst{\ref{inst31}}
\and N.~Zotov\inst{\ref{inst87}}
\and M.~E.~Zucker\inst{\ref{inst20}}
\and J.~Zweizig\inst{\ref{inst1}}
\end{spacing}
}
}

\institute{LIGO - California Institute of Technology, Pasadena, CA  91125, USA\label{inst1}
\and
California State University Fullerton, Fullerton CA 92831 USA\label{inst2}
\and
SUPA, University of Glasgow, Glasgow, G12 8QQ, United Kingdom\label{inst3}
\and
Laboratoire d'Annecy-le-Vieux de Physique des Particules (LAPP), Universit\'e de Savoie, CNRS/IN2P3, F-74941 Annecy-Le-Vieux, France\label{inst4}
\and
INFN, Sezione di Napoli $^a$; Universit\`a di Napoli 'Federico II'$^b$ Complesso Universitario di Monte S.Angelo, I-80126 Napoli; Universit\`a di Salerno, Fisciano, I-84084 Salerno$^c$, Italy\label{inst5}
\and
LIGO - Livingston Observatory, Livingston, LA  70754, USA\label{inst6}
\and
Albert-Einstein-Institut, Max-Planck-Institut f\"ur Gravitationsphysik, D-30167 Hannover, Germany\label{inst7}
\and
Leibniz Universit\"at Hannover, D-30167 Hannover, Germany\label{inst8}
\and
Nikhef, Science Park, Amsterdam, the Netherlands$^a$; VU University Amsterdam, De Boelelaan 1081, 1081 HV Amsterdam, the Netherlands$^b\dagger$\label{inst9}
\and
National Astronomical Observatory of Japan, Tokyo  181-8588, Japan\label{inst10}
\and
University of Wisconsin--Milwaukee, Milwaukee, WI  53201, USA\label{inst11}
\and
University of Florida, Gainesville, FL  32611, USA\label{inst12}
\and
University of Birmingham, Birmingham, B15 2TT, United Kingdom\label{inst13}
\and
INFN, Sezione di Roma$^a$; Universit\`a 'La Sapienza'$^b$, I-00185 Roma, Italy\label{inst14}
\and
LIGO - Hanford Observatory, Richland, WA  99352, USA\label{inst15}
\and
Albert-Einstein-Institut, Max-Planck-Institut f\"ur Gravitationsphysik, D-14476 Golm, Germany\label{inst16}
\and
Montana State University, Bozeman, MT 59717, USA\label{inst17}
\and
European Gravitational Observatory (EGO), I-56021 Cascina (PI), Italy\label{inst18}
\and
Syracuse University, Syracuse, NY  13244, USA\label{inst19}
\and
LIGO - Massachusetts Institute of Technology, Cambridge, MA 02139, USA\label{inst20}
\and
Laboratoire AstroParticule et Cosmologie (APC) Universit\'e Paris Diderot, CNRS: IN2P3, CEA: DSM/IRFU, Observatoire de Paris, 10 rue A.Domon et L.Duquet, 75013 Paris - France\label{inst21}
\and
Columbia University, New York, NY  10027, USA\label{inst22}
\and
INFN, Sezione di Pisa$^a$; Universit\`a di Pisa$^b$; I-56127 Pisa; Universit\`a di Siena, I-53100 Siena$^c$, Italy\label{inst23}
\and
Stanford University, Stanford, CA  94305, USA\label{inst24}
\and
IM-PAN 00-956 Warsaw$^a$; Astronomical Observatory Warsaw University 00-478 Warsaw$^b$; CAMK-PAN 00-716 Warsaw$^c$; Bia{\l}ystok University 15-424 Bia{\l}ystok$^d$; IPJ 05-400 \'Swierk-Otwock$^e$; Institute of Astronomy 65-265 Zielona G\'ora$^f$,  Poland\label{inst25}
\and
The University of Texas at Brownsville and Texas Southmost College, Brownsville, TX  78520, USA\label{inst26}
\and
San Jose State University, San Jose, CA 95192, USA\label{inst27}
\and
Moscow State University, Moscow, 119992, Russia\label{inst28}
\and
LAL, Universit\'e Paris-Sud, IN2P3/CNRS, F-91898 Orsay$^a$; ESPCI, CNRS,  F-75005 Paris$^b$, France\label{inst29}
\and
NASA/Goddard Space Flight Center, Greenbelt, MD  20771, USA\label{inst30}
\and
University of Western Australia, Crawley, WA 6009, Australia\label{inst31}
\and
The Pennsylvania State University, University Park, PA  16802, USA\label{inst32}
\and
Universit\'e Nice-Sophia-Antipolis, CNRS, Observatoire de la C\^ote d'Azur, F-06304 Nice$^a$; Institut de Physique de Rennes, CNRS, Universit\'e de Rennes 1, 35042 Rennes$^b$, France\label{inst33}
\and
Laboratoire des Mat\'eriaux Avanc\'es (LMA), IN2P3/CNRS, F-69622 Villeurbanne, Lyon, France\label{inst34}
\and
Washington State University, Pullman, WA 99164, USA\label{inst35}
\and
INFN, Sezione di Perugia$^a$; Universit\`a di Perugia$^b$, I-06123 Perugia,Italy\label{inst36}
\and
INFN, Sezione di Firenze, I-50019 Sesto Fiorentino$^a$; Universit\`a degli Studi di Urbino 'Carlo Bo', I-61029 Urbino$^b$, Italy\label{inst37}
\and
University of Oregon, Eugene, OR  97403, USA\label{inst38}
\and
Laboratoire Kastler Brossel, ENS, CNRS, UPMC, Universit\'e Pierre et Marie Curie, 4 Place Jussieu, F-75005 Paris, France\label{inst39}
\and
University of Maryland, College Park, MD 20742 USA\label{inst40}
\and
University of Massachusetts - Amherst, Amherst, MA 01003, USA\label{inst41}
\and
Canadian Institute for Theoretical Astrophysics, University of Toronto, Toronto, Ontario, M5S 3H8, Canada\label{inst42}
\and
Tsinghua University, Beijing 100084 China\label{inst43}
\and
University of Michigan, Ann Arbor, MI  48109, USA\label{inst44}
\and
Louisiana State University, Baton Rouge, LA  70803, USA\label{inst45}
\and
The University of Mississippi, University, MS 38677, USA\label{inst46}
\and
Charles Sturt University, Wagga Wagga, NSW 2678, Australia\label{inst47}
\and
Caltech-CaRT, Pasadena, CA  91125, USA\label{inst48}
\and
INFN, Sezione di Genova;  I-16146  Genova, Italy\label{inst49}
\and
Pusan National University, Busan 609-735, Korea\label{inst50}
\and
Australian National University, Canberra, ACT 0200, Australia\label{inst51}
\and
Carleton College, Northfield, MN  55057, USA\label{inst52}
\and
The University of Melbourne, Parkville, VIC 3010, Australia\label{inst53}
\and
Cardiff University, Cardiff, CF24 3AA, United Kingdom\label{inst54}
\and
INFN, Sezione di Roma Tor Vergata$^a$; Universit\`a di Roma Tor Vergata, I-00133 Roma$^b$; Universit\`a dell'Aquila, I-67100 L'Aquila$^c$, Italy\label{inst55}
\newpage
\and
University of Salerno, I-84084 Fisciano (Salerno), Italy and INFN (Sezione di Napoli), Italy\label{inst56}
\and
The University of Sheffield, Sheffield S10 2TN, United Kingdom\label{inst57}
\and
RMKI, H-1121 Budapest, Konkoly Thege Mikl\'os \'ut 29-33, Hungary$^dagger$\label{inst58}
\and
INFN, Gruppo Collegato di Trento$^a$ and Universit\`a di Trento$^b$,  I-38050 Povo, Trento, Italy;   INFN, Sezione di Padova$^c$ and Universit\`a di Padova$^d$, I-35131 Padova, Italy\label{inst59}
\and
Inter-University Centre for Astronomy and Astrophysics, Pune - 411007, India\label{inst60}
\and
California Institute of Technology, Pasadena, CA  91125, USA\label{inst61}
\and
Northwestern University, Evanston, IL  60208, USA\label{inst62}
\and
University of Cambridge, Cambridge, CB2 1TN, United Kingdom\label{inst63}
\and
The University of Texas at Austin, Austin, TX 78712, USA\label{inst64}
\and
E\"otv\"os Lor\'and University, Budapest, 1117 Hungary\label{inst65}
\and
University of Szeged, 6720 Szeged, D\'om t\'er 9, Hungary\label{inst66}
\and
Universitat de les Illes Balears, E-07122 Palma de Mallorca, Spain\label{inst67}
\and
Rutherford Appleton Laboratory, HSIC, Chilton, Didcot, Oxon OX11 0QX United Kingdom\label{inst68}
\and
Embry-Riddle Aeronautical University, Prescott, AZ   86301 USA\label{inst69}
\and
National Institute for Mathematical Sciences, Daejeon 305-390, Korea\label{inst70}
\and
Perimeter Institute for Theoretical Physics, Ontario, N2L 2Y5, Canada\label{inst71}
\and
University of New Hampshire, Durham, NH 03824, USA\label{inst72}
\and
University of Adelaide, Adelaide, SA 5005, Australia\label{inst73}
\and
University of Southampton, Southampton, SO17 1BJ, United Kingdom\label{inst74}
\and
University of Minnesota, Minneapolis, MN 55455, USA\label{inst75}
\and
Korea Institute of Science and Technology Information, Daejeon 305-806, Korea\label{inst76}
\and
Hobart and William Smith Colleges, Geneva, NY  14456, USA\label{inst77}
\and
Institute of Applied Physics, Nizhny Novgorod, 603950, Russia\label{inst78}
\and
Lund Observatory, Box 43, SE-221 00, Lund, Sweden\label{inst79}
\and
Hanyang University, Seoul 133-791, Korea\label{inst80}
\and
Seoul National University, Seoul 151-742, Korea\label{inst81}
\and
University of Strathclyde, Glasgow, G1 1XQ, United Kingdom\label{inst82}
\and
Southern University and A\&M College, Baton Rouge, LA  70813, USA\label{inst83}
\and
University of Rochester, Rochester, NY  14627, USA\label{inst84}
\and
Rochester Institute of Technology, Rochester, NY  14623, USA\label{inst85}
\and
University of Sannio at Benevento, I-82100 Benevento, Italy and INFN (Sezione di Napoli), Italy\label{inst86}
\and
Louisiana Tech University, Ruston, LA  71272, USA\label{inst87}
\and
McNeese State University, Lake Charles, LA 70609 USA\label{inst88}
\and
Andrews University, Berrien Springs, MI 49104 USA\label{inst89}
\and
Trinity University, San Antonio, TX  78212, USA\label{inst90}
\and
Southeastern Louisiana University, Hammond, LA  70402, USA\label{inst91}
}

\date{\today}
\abstract{}
\titlerunning{First Low-Latency LIGO+Virgo Search for Binary Inspirals and Their EM Counterparts}
\authorrunning{LSC and Virgo}
\thispagestyle{empty}
\maketitle

\begin{onecolumn}
\begin{center}
{\bf ABSTRACT}
\end{center}
\vspace{12 pt}
\noindent{\it Aims.} The detection and measurement of gravitational-waves from coalescing
neutron-star binary systems is an important science goal for ground-based
gravitational-wave detectors.  In addition to emitting gravitational-waves at
frequencies that span the most sensitive bands of the LIGO and Virgo detectors,
these sources are also amongst the most likely to produce an electromagnetic
counterpart to the gravitational-wave emission.  A joint detection of the
gravitational-wave and electromagnetic signals would provide a powerful new
probe for astronomy.

\noindent{\it Methods.} During the period between September 19 and October 20, 2010, the first 
low-latency search for gravitational-waves from binary inspirals in LIGO and
Virgo data was conducted.  The resulting triggers were sent to electromagnetic
observatories for followup.  We describe the generation and processing of the 
low-latency gravitational-wave triggers.  The results of the electromagnetic
image analysis will be described elsewhere.

\noindent{\it Results.} Over the course of the science run, three gravitational-wave triggers passed
all of the low-latency selection cuts.  Of these, one was followed up by several of our
observational partners.  Analysis of the gravitational-wave data leads to an
estimated false alarm rate of once every 6.4 days, falling far short of the
requirement for a detection based solely on gravitational-wave data.
\end{onecolumn}

\vspace{12 pt}
\noindent{\bf Key words.} Gravitational Waves -- Methods: observational
\begin{twocolumn}

\section{Introduction}
The direct detection and measurement of gravitational-waves from coalescing
neutron-star and black-hole binaries is a high-priority science goal for
ground-based gravitational-wave detectors. Event rate estimates suggest $\sim
1/100$\,y detectable by the initial LIGO--Virgo network rising to $\sim
50/$y for the advanced LIGO--Virgo detector network~\citep{ratesdoc}. 
Coalescing compact binary systems containing at least one neutron star may also
produce an electromagnetic counterpart to the gravitational-wave emission.
Observation of both the gravitational and electromagnetic emission will allow
astronomers to develop a complete picture of these energetic astronomical events. 

Indeed there is a growing body of evidence that most short, hard $\gamma$-ray bursts (SHGRBs) are the result of binary neutron star or neutron star--black
hole
coalescence~\citep{2006ApJ...638..354B,2005Natur.438..988B,2005Natur.437..855V,2009ApJ...690..231B,2006ApJ...650..281N}.  
SHGRBs are known to emit electromagnetic radiation across the
spectrum:  From prompt, high energy emission that lasts $\sim$ minutes, to optical
afterglows that decay on the order of hours and even radio emission that decays
on the scale of weeks~\citep{Nakar:2011cw}. Coincident electromagnetic and gravitational observation of a SHGRB could confirm that the
central engine of the $\gamma$-ray burst is indeed a coalescing compact binary. 

If the $\gamma$-ray emission in SHGRBs is beamed, and the central engine is a compact binary merger, there will be many compact 
binary gravitational-wave events for which no $\gamma$-rays are observed by astronomers. On the other hand, optical afterglows may 
be observable further off-axis than the $\gamma$-rays albeit somewhat dimmer. There are other proposals for electromagnetic emission 
from binary neutron-star mergers including supernova-like emission due to
the radioactive decay of heavy elements in the ejecta from the
merger~\citep{Li:1998bw,Metzger:2010sy}.  This mechanism has been dubbed a
``kilonova" since the luminosity peaks a thousand times higher than a typical
nova.  Kilonova emission is predicted to be isotropic and peaks roughly a day after merger.
Current blind optical transient searches, such as that being undertaken by the
Palomar Transient Factory, are only expected to find one of these events per
year if the neutron binary coalescence rate is at the optimistic end of current
estimates and that number increases substantially for the LSST~\citep{Metzger:2010sy}. 
Since the network of gravitational-wave detectors is essentially omni-directional, there is a strong case to undertake an observing campaign in which compact-binary merger candidates, identified in gravitational-wave data, are rapidly followed up with electromagnetic observations.  

In this paper we report on the first low-latency search for gravitational-waves
from compact binary coalescence (CBC) in which detection candidates were
followed up with optical observations.  The search covered the period of time between
September 19 and October 20, 2010, when LIGO detectors at Hanford, WA (H1),
Livingston, LA (L1) and the Virgo detector in Cascina, Italy (V1) were jointly
taking data.  This time was contained within LIGO's sixth science (S6) which
began on July 7, 2009 and Virgo's third science run which began on August 11,
2010.  The group of astronomical partners prepared to make followup
observations included the Liverpool telescope, LOFAR, the Palomar Transient
Factory, Pi of the Sky, QUEST, ROTSE III, SkyMapper, Swift, TAROT and the Zadko
telescope.  
A companion paper~\citep{2011arXiv1109.3498T} describes more generally the
process of collecting gravitational-wave triggers from a variety of transient
searches, the human monitoring of the process and the production of
telescope tilings.

The search resulted in three triggers that passed all of the
selection cuts.  The first occurred during a testing period and no alert was
sent out.  The second trigger was sent out to our astronomical partners and
images were taken by several of them.  An alert was issued for the third
trigger, though its location was too close to the sun to be observed.

This paper is organized as follows:  In Sect.~II we provide a description of the
analysis pipeline that produced triggers and localized them in the sky.  In
Sect.~III we present a performance comparison between the low-latency pipeline
used to generate astronomical alerts and the standard trigger generator used in
offline searches for compact binary inspirals.  Sect.~IV presents an overview of
the run and the details of the triggers that were produced.

\section{The Pipeline}
The major components of the analysis pipeline are shown in Fig.~\ref{pipeline}
and described in this section. Gravitational-wave data from the three
participating detectors was first transfered to the Virgo site where it was
processed by the \emph{Multi-Band Template Analysis} (MBTA) to look for CBC signals. Interesting coincident triggers identified by MBTA were submitted to the 
\emph{Gravitational-wave Candidate Event Database} (\gracedb). A realtime alert
system (\lvalert) was used to communicate information about these events to listening
processes that automatically checked available information for possible
anomalous detector behavior and
estimated the location of the source on the sky. These automated steps completed
within a few minutes after data acquisition.

Further processing was needed before an alert was sent to our astronomical partners: human monitors reviewed information about the event, consulted the detector control rooms, and examined the data quality using a number of higher latency tools. Telescope image tilings were generated simultaneously in preparation for a positive decision to follow-up a trigger. The entire
process took 20-40 minutes, with the largest latency incurred by
the human monitor step.  A histogram of the latency incurred before the trigger
is sent out for further processing is shown in Fig.~\ref{lathist}.  Details about the rest of the event processing
can be found in~\citep{2011arXiv1109.3498T}.  

\begin{figure}[!h]
  \centering
\includegraphics[width=8cm]{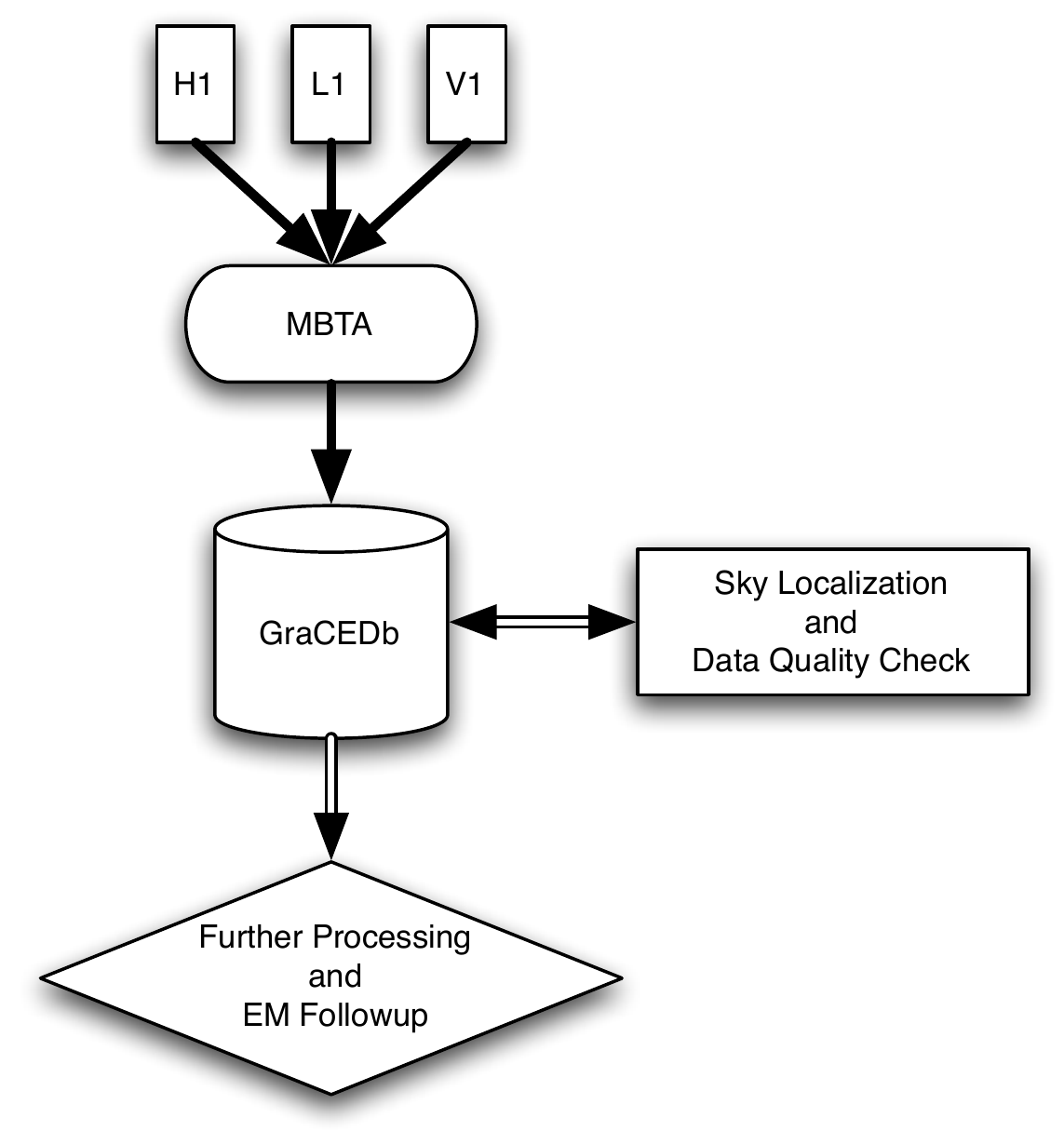}
\caption{An overview of the pipeline.  Data produced at each of the three
detector sites is transfered to a computer at Cascina where
triggers are produced by MBTA and sent to
GraCEDb for storage.  Upon receiving events, GraCEDb alerts the sky localization
and data quality check processes to begin and the results are then sent back to
GraCEDb.  If the triggers are localizable and of acceptable data quality then
they are sent out for further processing and possibly followup by an optical
telescope.  The double stroke connections in the diagram are provided by
LVAlert.}\label{pipeline} 
\end{figure}

\subsection{Trigger production with MBTA}
\subsubsection{Pipeline structure and parameters}
The Multi-Band Template Analysis (MBTA)~\citep{Beauville:2007kp} is a low-latency implementation of the
standard matched filter~\citep{wainstein:1962} that is commonly used to search
for gravitational-waves from compact binary coalescences (CBCs).  As such, it
relies on the accurate mathematical models of the expected signals to be used as search
templates.  Time domain templates with phase evolution accurate to second
post-Newtonian order were used in the search.  The search covered sources 
with component mass between $1-34\ M_{\odot}$ and total mass below 
$35\ M_{\odot}$. A fixed bank of templates 
constructed with a minimal match of 97\% was used to scan this parameter space. 
The template bank was constructed using a reference noise
power spectrum taken at a time when the detectors were performing well.
Given the detectors' typical noise spectra, the 
low-frequency cutoff of the templates was set to $50$\,Hz, thus keeping the 
computational cost of the search light while losing negligible
signal-to-noise ratio (SNR).
\par
An original feature of MBTA is that it divides the matched filter across two
frequency bands.  This results in two immediate benefits: (1) The phase of the signal is 
tracked over fewer cycles meaning sparser template banks are needed in each
frequency band and (2) a reduced sampling rate can be used for the lower 
frequency band, reducing the computational cost of the fast Fourier transforms 
involved in the filtering.  The full band signal-to-noise ratio $\rho$ is 
computed by coherently combining the matched filtering outputs from the two 
frequency bands. The boundary between the low and high frequency bands is 
chosen such that the SNR is roughly equally shared between the two bands. The 
boundary frequencies range from $110$\,Hz to $156$\,Hz, depending on the detector 
and on the mass ratio of the template bank of the individual MBTA processes.
\par
A trigger is registered when $\rho > 5.5$. 
Triggers are clustered across the template bank and across 
time: triggers less than $20$ms apart are recursively clustered under the loudest trigger. 
Clustered triggers are subjected to a $\chi^2$ test to check if the SNR
distribution across the 
two frequency bands is consistent with the expected signal. The discriminating power of such a 
2-band $\chi^2$ test is not as high as in typical implementations based on a larger number of 
frequency bands, but offers the advantage of having a negligible computational cost in the multi-band 
framework. The test can therefore be applied to all triggers at the single detector level. 
Triggers pass the $\chi^2$-test if
\begin{equation}
\chi^2 < 3\ (2 + 0.025\ \rho^2) \; .
\end{equation}
\par
Single detector triggers that pass the $\chi^2$ test are tested for
coincidence across detectors.  Triggers from two detectors $i$ and $j$ are 
considered coincident if their time and mass parameters match within expected
uncertainties. The mass coincidence criterion is
\begin{equation}
\left| {\cal M}_i - {\cal M}_j \right| < \left(\frac{0.05}{M_\odot}\right)\ \left( \frac{{\cal M}_i + 
{\cal M}_j}{2}\right)^2
\end{equation}
where the chirp mass is given by${\cal M}=(m_1m_2)^{3/5}(m_1+m_2)^{-1/5}$ in
terms of the component masses, $m_1$ and $m_2$.
The time coincidence criterion is
\begin{equation}
\left| t_i - t_j \right| < \Delta t_{ij}
\end{equation}
where the maximum allowed time delays $\Delta t_{H1L1} = 20$\,ms and $\Delta t_{H1V1} = 
\Delta t_{L1V1} 
= 40$\,ms account for both the time of flight of the signal from one detector
to another and for the experimental uncertainty in the arrival time measurement.
The arrival time is measured by performing a quadratic fit around the maximum of the match 
filtered signal. It is then extrapolated to 
the time when the gravitational-wave signal is at a reference
frequency chosen to minimize the statistical uncertainty on the measurement.
This has the important effect of reducing the background by allowing tighter
coincidence and improving the accuracy of position reconstruction by
triangulation. It has been shown elsewhere~\citep{Acernese:2007zza} that the
optimal reference frequency is
approximately that frequency where the detectors are most sensitive, although
it depends somewhat on the mass of the source. A detailed study of simulated
signals in S6/VSR3 noise resulted in the empirical parametrization  $f_{\mathrm{reference}}=
[170 - {\cal M}(5.1/M_\odot)]$\,Hz.
\par
Triple detector coincidences are identified as a pair of H1L1 and H1V1 coincidences sharing 
the same H1 trigger. Although the pipeline identifies both double and triple coincidences, 
only the latter were submitted to \gracedb, in line with the intention to focus on 
remarkable events, for which a sky location could be extracted by simple triangulation. 
Each triple has associated with it a combined SNR $\rho_c$ defined by 
\begin{equation}
\rho_c^2 = \sum_{j \in \{\mathrm{H1},\mathrm{L1},\mathrm{V1}\}} \rho_{j}^2
\end{equation}
and a false alarm rate (FAR).  Triggers with a smaller FAR, i.e. a larger
$\rho_c$, are more likely to be gravitational-waves.  The non-stationary nature
of the background means that a simple mapping from a given $\rho_c$ to a
corresponding FAR does not exist.  Instead, the FAR must be explicitly
calculated for each trigger.
\par
The FAR is estimated as follows. Let $T_i$ be the
analysis time needed to accumulate the last 100 triggers in detector $i$. Let
$N_{\mathrm louder}$ be the number of mass-coincidences that can be formed from
the last 100 triggers in each detector for which $\rho_c$ is greater than the
combined SNR of the observed coincidence. Then the expected rate of coincident
triggers arising from background may be estimated as the product of the
individual detector rates multiplied by a factor that accounts for the
coincidence windows in time and mass. In particular, we define the FAR
associated with a triple-coincident trigger to be
\begin{equation}
{\mathrm FAR} = \alpha\ N_{\mathrm{louder}} \frac{4\ \Delta t_{H1L1}\ \Delta
t_{H1V1}}{T_{H1}\ T_{L1}\ T_{V1}}
\end{equation}
where $\alpha$ is 
an empirical factor with a value of $2$ tuned to adjust the estimated FAR to the
average observed rate of such triple-coincident triggers.
\par
The low latency search is dependent on the short-order
availability of the strain time series $h(t)$ measured by
the detectors. Tools to reconstruct $h(t)$ with very low latency
have been developed over the last few years, and had
reached a mature state by the time of the S6/VSR3 runs. Calibration 
accuracies better than 15\% on the amplitude of $h(t)$ were typically 
achieved for all detectors, which is quite sufficient for the purposes of our
search. 
\par
For each detector, the $h(t)$ channel is produced at the
experimental site computing center. The H1 and L1 $h(t)$
data are then transfered to the Virgo site computing center,
where all the MBTA processing takes place. To minimize the latency, 
the communication of input/output data between the different processes 
involves no files on disk. It relies instead on the use of shared memories and
of a TCP-IP based communication protocol developed
for the Virgo data acquisition system. The whole set of processes
ran on six computers.
\subsection{GraCEDb and LVAlert}
The storage and communication capabilities of the pipeline in
Fig.~\ref{pipeline} are provided by the
Gravitational-wave Candidate Event Database (\gracedb) and the LValert messaging
system.  The purpose of these technologies is to provide a continuously running
system to ingest, archive and respond to gravitational-wave
triggers.  Communication with individual telescopes is handled at a later stage
and uses whatever protocol is appropriate for the particular telescope.

The \gracedb\ service stands behind an Apache~\citep{Apache} server and is built
on Django~\citep{Django}, a
command line client that uses an
HTTP/ReST~\citep{Fielding_2000_Architectural-Styles,Fielding02principleddesign} interface to the server and
authenticates with X509~\citep{X509} certificates to a MySQL~\citep{MySQL}
database back-end. A
command-line client allows easy automation of event submission to \gracedb.
The prototype system used during S6/VSR3 was capable of ingesting triggers from
MBTA and a number of other search pipelines. The raw trigger information was
stored in an easily accessible archive and the most relevant information about
the trigger, such as the time and significance, were ingested into the
database.  Upon successful ingestion, the trigger is given a unique identifier
that is returned to the submitter. An alert is then sent out via LVAlert.

LVAlert is a communication client built on XMPP~\citep{XMPP} technology and the PubSub
extension~\citep{PUBSUB}. The system allows users to create nodes to which
information can be published; users subscribe to nodes from which they want to
receive that information. In the context of the current search, a node was set
up for sending out alerts about MBTA triggers.  A listener client is also
provided that waits for alerts from the nodes to which the user is subscribed.
By default, the listener simply prints any information it receives to the
\texttt{stdout}, but it also allows users to develop plugins which can take
action in response to the alerts. 
This is the mechanism for launching the data quality and sky localization
jobs in Fig.~\ref{pipeline}.

\begin{figure}[!h]
  \centering
\includegraphics[width=8cm]{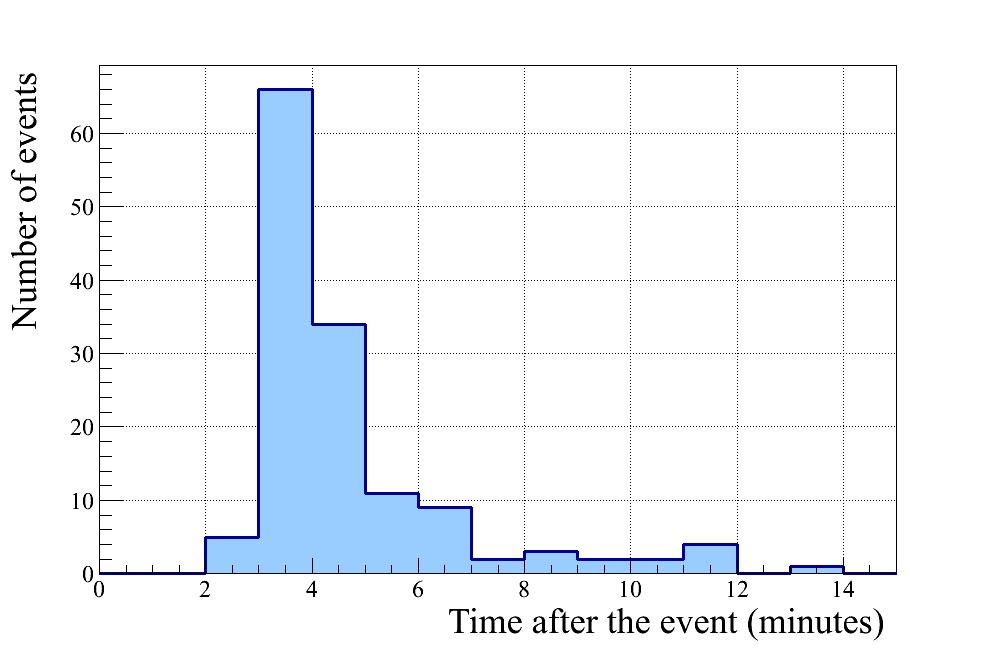}
\caption{Total latency of the automatic processes during the S6/VSR3 run.  The x-axis is
the difference between the time at which the trigger was available for human monitoring and the
trigger's time. This includes an average of $63$\,s for all the data to be available at the Virgo site
computing center and an average of $142$\,s for the trigger to be submitted to GraCEDB. This does not 
include the time for the human monitoring ($\sim$20-40 minutes)
that takes place before an event is sent out for electromagnetic followup.}\label{lathist} 
\end{figure}

\subsection{Event processing}
While MBTA is responsible for producing coincident triggers, there remains
further processing to determine whether or not the triggers are suitable for
external followups.  In particular, this processing consists of performing sky
localization, checking the quality of the data and determining whether or not
the event is a known hardware injection, as described below.  The mechanism for
performing these tasks is an LVAlert listener that responds to alerts sent out
by GraCEDb in response to new MBTA triggers.

The sky localization procedure proceeds by computing, as a function of location
on the sky for a pre-determined grid, the errors in timing and
amplitude.  Probabilities are assigned by comparing these measured
quantities with distributions of these quantities obtained from simulated
signals.  More specifically, the measure of timing accuracy used is
\begin{equation}  \label{deltaTrms}
\Delta t_{\mathrm{rss}}(\alpha,\delta) = \sqrt{\sum_{i\neq j} {[ (t_i(\alpha,\delta) -
\hat{t}_i) - (t_j(\alpha,\delta) - \hat{t}_j)]}^2},
\end{equation}
where $t_i(\alpha,\delta)$ is the geocentric arrival time\footnote{We assume that
gravitational-waves travel at the speed of light.} of a source located at right
ascension $\alpha$ and declination $\delta$ in detector $i$ and $\hat{t}_i$ is
the measured arrival time.  To improve sky localization performance, the time of 
arrival is taken to be the time the signal crosses the reference
frequency~\citep{Acernese:2007zza} of $140$Hz.  The amplitude consistency parameter is given by
\begin{equation}
\Delta Q_{\rm rss}(\alpha,\delta) = \sqrt{\sum_{i\neq j} \left(\frac{D_{{\rm eff}}^{(i)\,2} - 
D_{{\rm eff}}^{(j)\,2}}{D_{{\rm eff}}^{(i)\,2} + D_{{\rm eff}}^{(j)\,2}} -
\frac{Q^{(i)\,2} - Q^{(j)\,2}}{Q^{(i)\,2} + Q^{(j)\,2}}\right)^2},
\end{equation}
where the $(i)$ superscripts label the detector, $D_{\rm eff}$ is the {\it
effective distance} and we have suppressed the $(\alpha,\delta)$ of every
quantity on the right hand side of the equation.  Analytically, the effective
distance is defined by 
\begin{equation}\label{deffdef}
D_{\rm eff} = D\left[F_{+}^2\left(\frac{1+\cos^2\iota}{2}
\right)^2 + F_{\times}^2\cos^2\iota\right]^{-1/2},\label{deffdef}
\end{equation}
where $D$ is the actual distance to the source, $F_{+,\times} = 
F_{+,\times}(\alpha,\delta,\psi)$ are the response functions of the detector
(that depend on the polarization angle, $\psi$) and $\iota$ is the inclination 
angle of the source, and the $(i)$ superscript is understood.
In practice, the matched filtering procedure only provides
a measurement $D_{\rm eff}$ that is not typically invertible to obtain $D$,
$F_{+,\times}$, etc.  We introduce the quantity (suppressing the $(i)$
superscript once again)
\begin{equation}
Q^2 \equiv \frac{1}{F_{+}^2(\theta,\phi,\psi=0) +
F_{\times}^2(\theta,\phi,\psi=0)},
\end{equation}
to provide an {\emph ad hoc} measure for determining the consistency of the
amplitudes measured in each detector with a particular sky location.  In
practice, the use of this quantity improves the sky localization accuracy by roughly
a factor of two, meaning it helps to break the mirror-image degeneracy inherent
in a three-detector network when only triangulation is used to locate the source
on the sky.  Values of $\Delta t_{\mathrm{rss}}(\alpha,\delta)$ and
$\Delta Q_{\rm rss}(\alpha,\delta)$ are assigned probabilities by comparing
their values to those of a distribution of simulated signals.  More
specifically, simulated signals are placed into detector data and probability
distributions are computed from the values of  $\Delta
t_{\mathrm{rss}}(\alpha_{\rm true},\delta_{\rm true})$ and
$\Delta Q_{\rm rss}(\alpha_{\rm true},\delta_{\rm true})$, where the subscript
indicates that the quantity is evaluated at the true location of the source in the sky.  
Since there is negligible
correlation between the timing and amplitude measures, the normalized product of
their values yields the probability distribution over the sky, i.e., the skymap.

As a demonstration of the performance of the sky localization routine, 10122
simulated signals were injected into data taken from the $6^{\mathrm th}$ week
of S6/VSR2.  The locations on the sky and distances were given by a galaxy
catalog~\citep{LIGOS3S4Galaxies}.  More details about these injections can be found in Sect.~\ref{comp}.  
The SNR distribution of the population of injections recovered by the pipeline 
is given in Fig.~\ref{snrdist}.  We use two figures of merit to assess sky
localization performance.  The first is a quantity we call {\it searched area}.
It is the area on the sky contained in the pixels that are assigned a
probability greater than the probability assigned to the pixel containing the
true source location.  In other words, it is the smallest area on the sky that 
one would be required to image to ensure the true source location is imaged.  
A complimentary figure of merit is the angular distance between the true source
location and the most probable location identified by the sky localization 
routine.  These figures of merit are shown as cumulative histograms in
Fig.~\ref{skyperf}.  To improve the performance of the sky localization
routine, a galaxy catalog is used to guide the pointing in the following way:
The skymap is weighted by the fraction of the cumulative blue luminosity (used
in this context as a proxy for star formation rates) in each pixel out to the 
smallest effective distance measured by any of the
detectors~\citep{2011arXiv1109.3498T}.  The effect of
using a galaxy catalog is evident in the figures of merit shown in
the upper curves in Fig.~\ref{skyperf}.  Focusing on the lower curves, the angular
distance figure of merit shows that for 60\% of the injections the maximum
likelihood point on the skymap is $\lesssim 5$\,degrees away from the true
source location, whereas the searched area is a much larger region on the
sky.  This indicates that although the maxima of the skymaps are close to the
true source locations, they are quite broad.  Application
of the galaxy catalog does not do much to the angular distance histogram, but it
eliminates large regions of the sky, effectively making the broad peaks much
sharper.  In Fig.~\ref{skyperfSNR} the median searched area is plotted as a function of
combined SNR.  The upper panel depicts the results without the aid of  the
galaxy catalog, while the lower panel includes those improvements.  A least
squares fit of the median searched areas is given by the red lines.  The fit
is to a functional form $a_0 + a_{-1}/\rho + a_{-2}/\rho^2$.
The results that make use of the galaxy catalog should be understood
as upper limits on the improvements a galaxy catalog can bring.  As described in
Sect.~\ref{comp}, the injection distribution was such that injections were more
likely to come from galaxies with larger blue light luminosity.  Although the
choice to use the blue light luminosity is well motivated, it does not
include, for example, binaries arising in elliptical galaxies, or the
possibility that binaries follow some other property of their host galaxies.
Better astrophysical priors are a subject we leave for future investigations.

\begin{figure}[!h]
  \centering
\includegraphics[width=8cm]{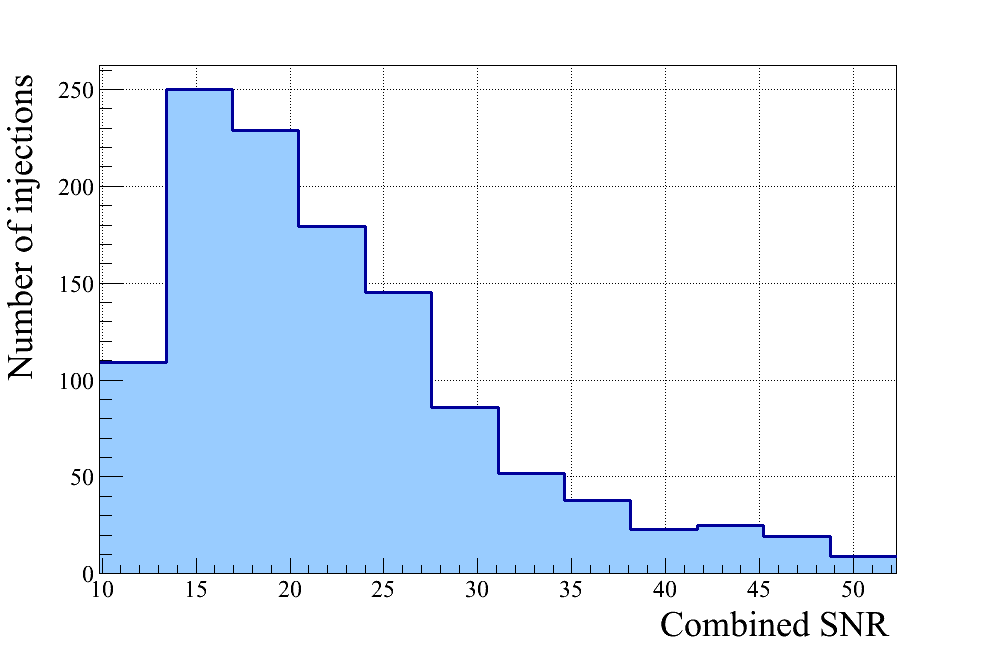}
\caption{The combined SNR distribution of the family of found injections used to
demonstrate the sky localization performance. Emphasis was placed on weaker
signals, where a first detection is more likely to arise.} \label{snrdist}
\end{figure}

\begin{figure}[!h]
  \centering
\includegraphics[width=8cm]{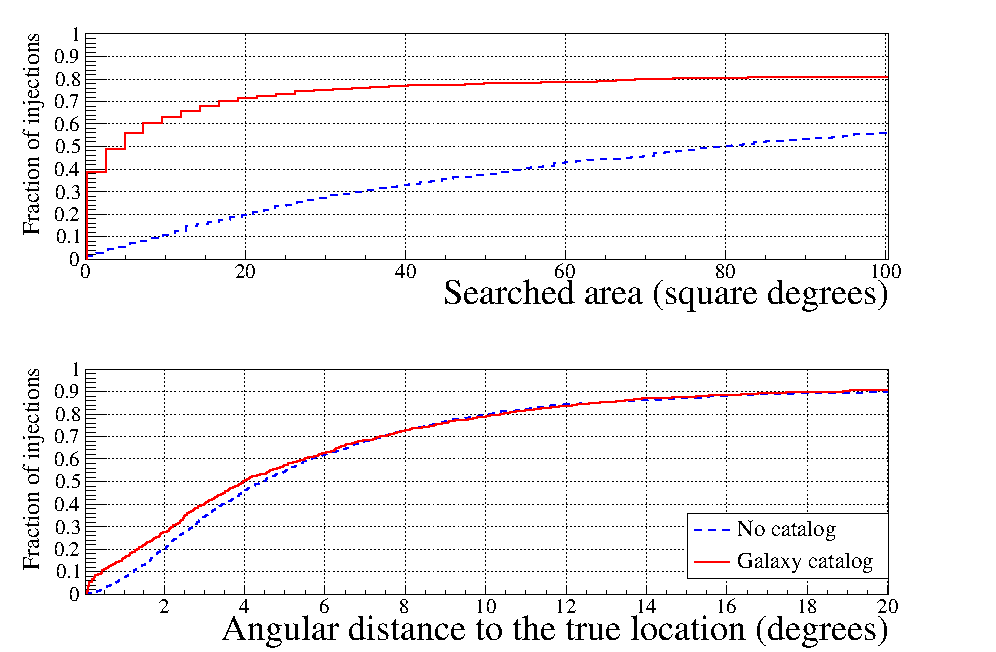}
\caption{Sky localization performance with and without
the use of a galaxy catalog.  The upper pane shows a cumulative
histogram of the searched area in square degrees.  The lower pane is a
cumulative histogram of the angular distance, in degrees, between the injected
location and the maximum likelihood recovered location. In both plots the red
solid line is the performance with the aid of the galaxy catalog and the blue
dotted line is the performance without the galaxy catalog.} \label{skyperf}
\end{figure}

\begin{figure}[!h]
  \centering
\includegraphics[width=8cm]{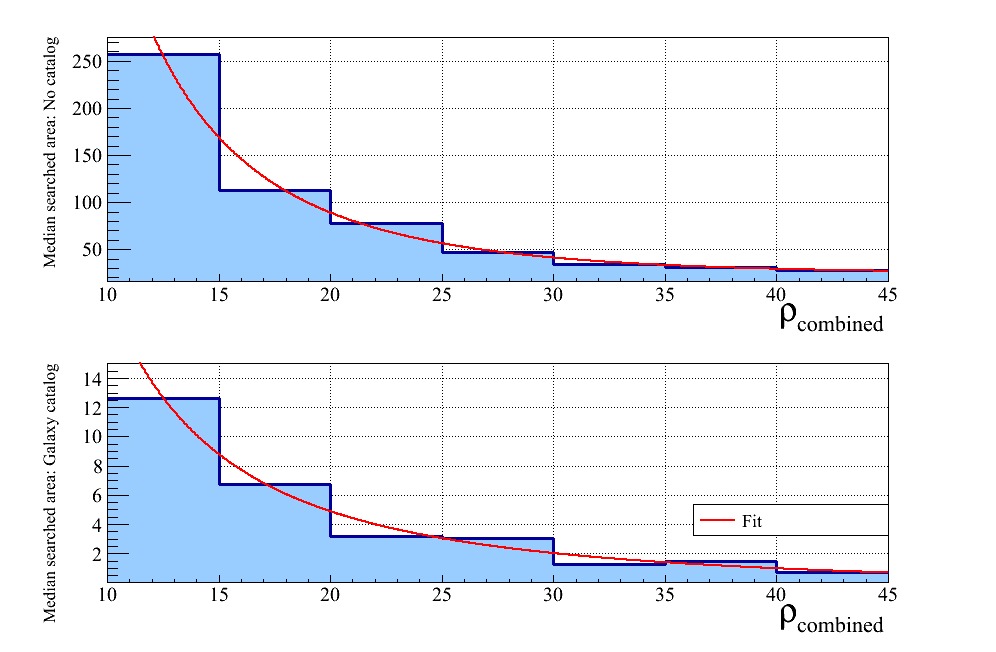}
\caption{Sky localization performance as a function of combined signal-to-noise
ratio with and without the use of a galaxy catalog (bottom and top panel,
respectively).  The blue bars indicated the median searched area, in square
degrees, in each bin and
the red lines depict a fit to these values.} \label{skyperfSNR}
\end{figure}

Data quality plays an important role in offline searches, where detailed studies are performed
to select data free from obvious instrumental or environmental artifacts by using information in
auxiliary channels. While all the data quality information is not available with low latency,
online production of a significant number of data quality flags was implemented during S6/VSR3.
This allows for quick feedback on the current data quality, and to check the data at the
time of a trigger against a list of trusted data quality flags produced online. As new data quality
flags are developed and tested, the list is updated accordingly.

If an event passes all of our checks then it is sent out to LUMIN, where it is
then processed further and possibly sent out for electromagnetic observation.

\section{MBTA Performance and Validation}
In this section we look at the performance of MBTA.  There are typically two ways
in which performance is assessed: (1) through hardware injections, where the
mirrors of each detector are physically displaced to simulate the presence of a
signal and (2) software injections in which simulated signals are placed into
detector data.  We will first look at MBTA's performance on hardware
injections
and then provide a detailed comparison of its performance relative to the CBC
offline pipeline~\citep{LIGOS3S4all, LIGOS3S4Tuning,
Collaboration:2009tt, Abbott:2009qj, S5LowMassLV}, known as iHope, on a set of common software injections.

\subsection{Hardware injections}
Hardware injections are produced through the displacement of the mirror 
located at the end of one of the detector's arm. They are performed 
coherently in the three detectors, i.e. with injection time, amplitude and
phase chosen in each detector to be consistent with the location of the
source on the sky.  During the S6/VSR3 scientific
run, there were approximatively three hardware injections per 
day in each detector simulating the coalescence of a compact binary system, 
with a period of intensive injections between August 27, 2010 and September 3,
2010. For a number of reasons, not every injection was successful at every
detector.

Two families of non-spinning waveforms were used for the hardware injections.
The first family corresponds to analytic
inspiral-merger-ringdown waveforms based on the Effective One-Body (EOB)
model extended and tuned to match Numerical Relativity simulations of
binary black hole coalescences \citep{Buonanno:2007pf,Pan:2007nw}. 
The second family of waveforms corresponds to restricted parameterized 2PN inspiral 
waveform computed in the time domain \citep{Blanchet:1996wx,Arun:2004ff}.  
Among the hardware injections successfully performed in all
three detectors, 62\% are from the first family and 38\% are from the second. 

The parameters of the hardware injections were 
adjusted to appear in each detectors with an $\textrm{SNR} < 100$. 
The total masses of the simulated binary systems were distributed 
between $2.8\, M_\odot$ and $35\, M_\odot$, with the masses of the components 
between $1.4\, M_\odot$ and $35\, M_\odot$. The hardware injections were distributed between
$1$\,Mpc and $80$\,Mpc, and the sky locations were chosen randomly. Figure~\ref{Plot_HardInj_3}
shows the chirp mass distribution of all the hardware injections which were
successfully found by MBTA during the S6/VSR3 run.  

\begin{figure}[!h]
  \centering  
\includegraphics[width=8cm]{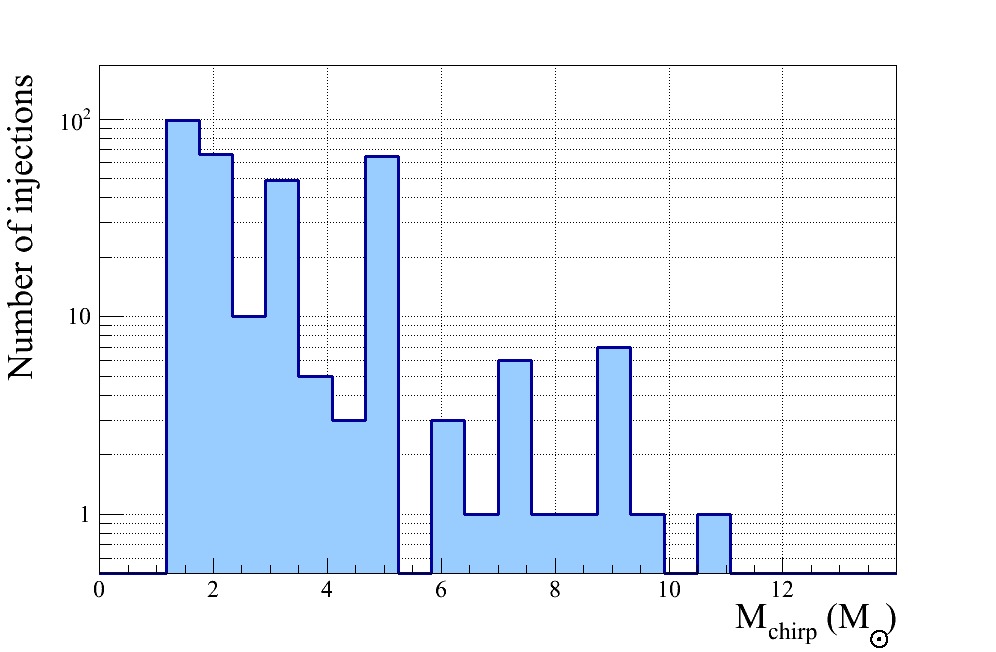}  
\caption{The chirp mass distribution of all the hardware injections found in a single detector 
by both MBTA and the offline iHope pipeline during the S6/VSR3 run. } \label{Plot_HardInj_3}  
\end{figure}

An injection is considered to be recovered if MBTA produced a trigger with an
end time within $20$\,ms of the end time of the injection. We
restrict our attention to injections occurring when all three detectors are
taking science quality data and all of the MBTA processes are up and running.
Our focus is further narrowed by the requirement that the data quality is
acceptable in $60$\,s preceding a trigger.  This is to ensure that the Fourier
transform required by the matched filter can be performed.  Given these caveats,
MBTA achieved a 93\% efficiency in detecting hardware injections in triple coincidence during the
S6/VSR3 run. Figure~\ref{Plot_HardInj_1} shows these recovered hardware 
injections and indicates the type of coincidence (double or triple) they were 
found in.  

\begin{figure}[!h]
  \centering 
\includegraphics[width=8cm]{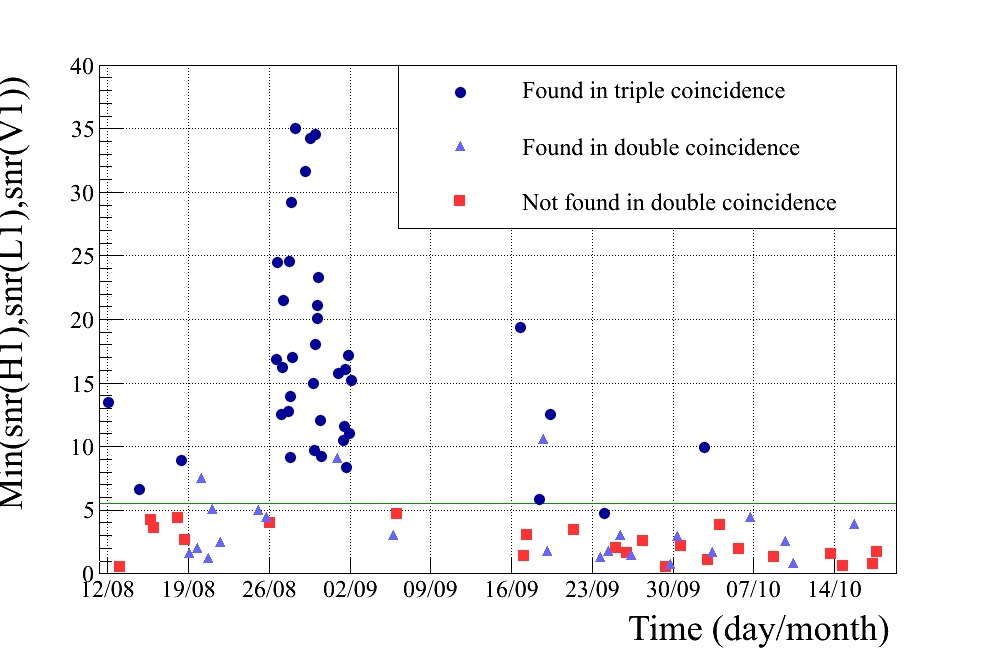} 
\caption{This plot shows all the successful hardware injection which took place 
coherently in the three detectors. 93\% of those which where above the 
threshold (expected $\textrm{SNR} \geq 5.5$) in each detector were detected in 
triple coincidence by MBTA.} \label{Plot_HardInj_1} 
\end{figure}

\begin{figure}
  \centering
\includegraphics[width=8cm]{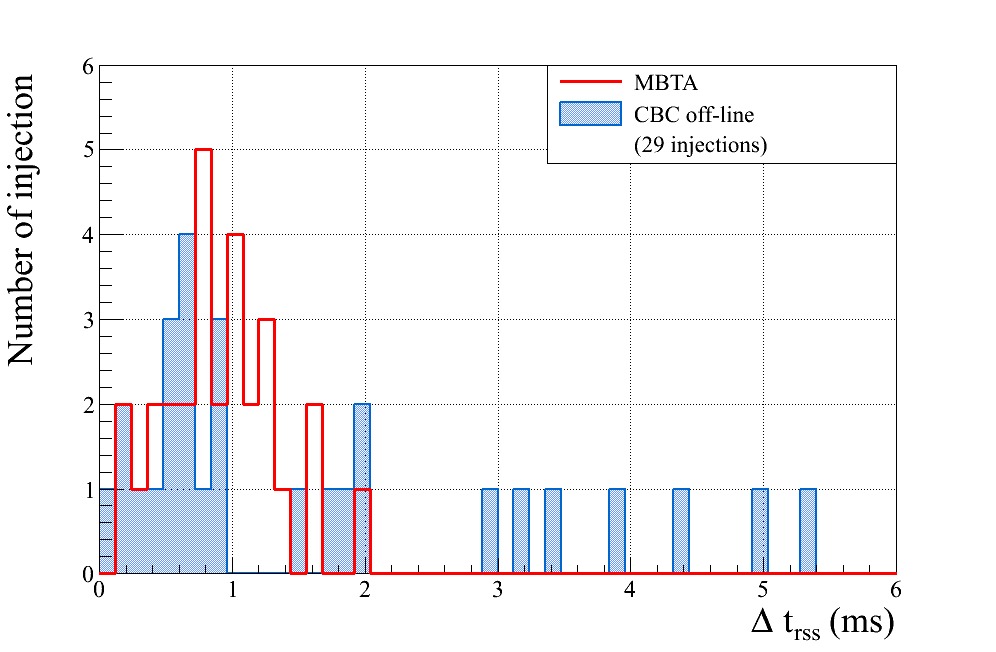} \includegraphics[width=8cm]{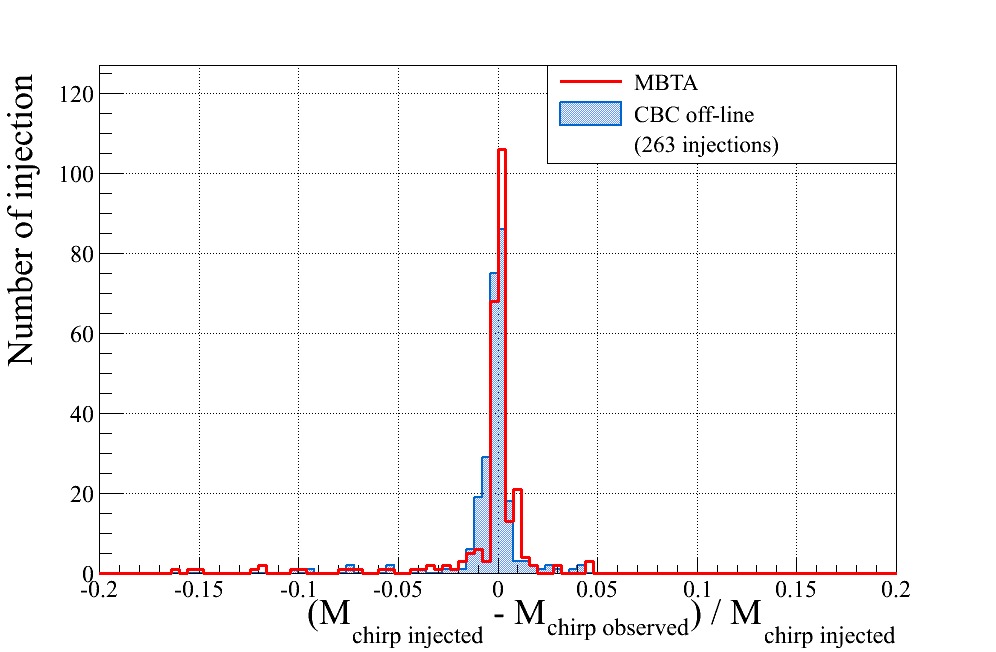}
 \includegraphics[width=8cm]{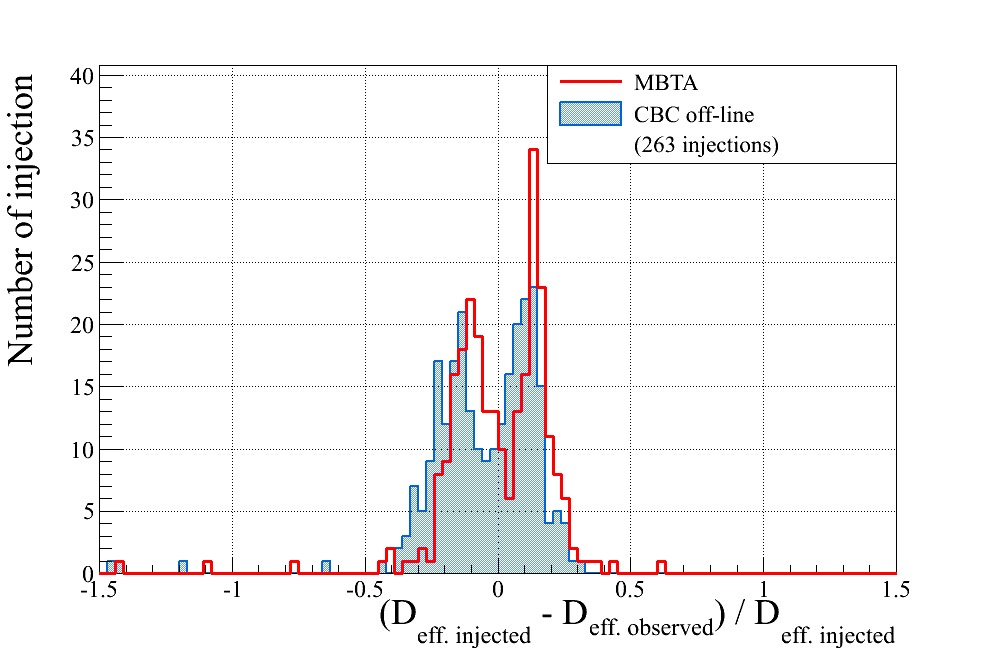}
\caption{These plots show the differences between the parameters of the 
hardware injections and those of the triggers found by both MBTA and the offline 
iHope pipeline. The first plot shows the timing accuracy parameter $\Delta t_{\mathrm{rss}}$
 of all triple coincidences. The second plot 
corresponds to the difference between the chirp mass of the binary injected and 
observed. The third one correspond to the difference between the effective distance of
the binary injected and reconstructed by MBTA. The second and third plots contain all the hardware
 injections observed in H1, L1 or V1.} \label{Plot_HardInj_2}
\end{figure}
An important step in the validation of MBTA is the comparison of the injected
signal parameters with those recovered by MBTA. Figure~\ref{Plot_HardInj_2} shows that
the parameters of the injected signals are in good agreement with those recovered by MBTA. 
The first plot of Fig.~\ref{Plot_HardInj_2} shows the timing accuracy parameter of the LIGO-Virgo
network, $\Delta t_{\mathrm{rss}}$, given by Eq.~\ref{deltaTrms}.
The second and third plots of Fig.~\ref{Plot_HardInj_2} show
the difference between the chirp
mass and effective distance of the binary injected and observed. From these three plots,
it is clear that the low-latency search provides parameter estimations with
comparable accuracy to that of the offline iHope pipeline.

\subsection{\label{comp}Software Injections}
Hardware injections are for obvious reasons impractical for large scale
performance studies.  Instead, it is preferable to  inject simulated signals
into data that is already on disk.  Here we will compare the performance of MBTA
with the existing offline CBC pipeline, known as iHope. Both
pipelines were run on a week of data containing simulated binary neutron star
and neutron star-black hole binary signals.  For this study a total of $10122$
software injections were performed and analyzed in multiple parallel runs.

Simulated waveforms were generated using the stationary phase approximation 
(SPA)~\citep{Droz:1999qx, thorne.k:1987,SathyaDhurandhar:1991} with the
amplitude expanded to Newtonian 
order and the phase expanded to second post-Newtonian order with the upper 
cut-off frequency at the Schwarzschild innermost stable circular orbit (ISCO). 
Their  locations were randomly  chosen from the Compact Binary Coalescence Galaxy 
Catalogue~\citep{LIGOS3S4Galaxies} in such a manner that the probability of choosing a 
galaxy is proportional to its blue light luminosity.  Injections were made out
to a distance of $40$\,Mpc which roughly corresponds to the 
limits of sensitivity of the LIGO-Virgo 
network during the observation period, for the considered sources. The mass of each binary component 
ranged from $1\, M_{\odot}$ to $15\, M_{\odot}$ with the constraint on the total 
mass of the binary to be between  $2\, M_{\odot}$ and  $20\, M_{\odot}$.  These
choices were made to focus on systems likely to contain at least one neutron
star, where an electromagnetic counterpart is expected.

After data quality vetoes were applied MBTA and iHope recovered 
different numbers of triple-coincident triggers, 736 and 859, respectively. 
Both pipelines recovered 709 identical injections in triple coincidence.
Among the rest of the MBTA triples, twenty five  were found by the 
offline CBC pipeline in double coincidence and the remaining two were 
completely missed. All these signals were near the threshold of detectability 
in one or more detectors.

\paragraph{Timing accuracy}
The primary goal of the low-latency pipeline is to send triggers out to the
astronomical community for electromagnetic followup observations. Hence,
localizing the GW candidate event on the sky is one of the essential
parts of the search.  Good timing accuracy for recovered injections is
essential for good sky localization.
Figure~\ref{fig:comp1} shows the normalized distributions of 
the timing accuracy parameter of the  LIGO-Virgo network, $\Delta
t_{\mathrm{rss}}$, given by (\ref{deltaTrms}). From this plot it is clear that, overall, MBTA's 
performance in recovering the arrival time of gravitational-wave is better or 
comparable to that of the offline search.  This is expected because MBTA uses a
quadratic fit to find the peak of the SNR time series whereas the offline
pipeline simply takes the maximum of the time series.

\begin{figure}[!h]
  \centering
\includegraphics[width=0.4\textwidth]{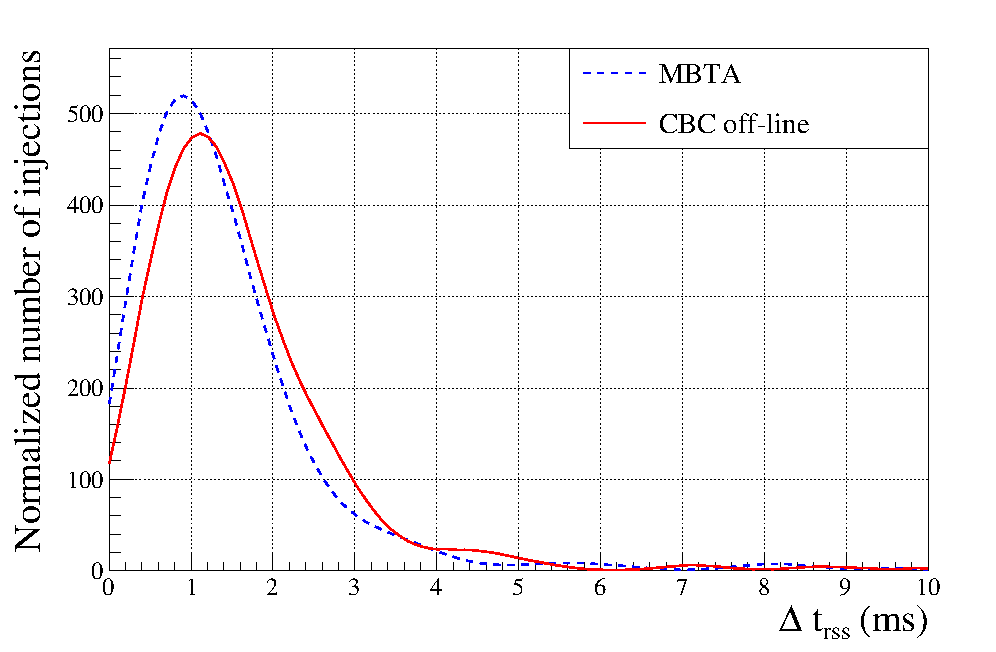}
  \caption{Normalized distribution of timing accuracy of triggers
    detected by MBTA and CBC offline analysis pipeline. MBTA shows
    slightly better  performance than the offline analysis.} \label{fig:comp1}
\end{figure}

\begin{figure}[!h]
  \centering
  \includegraphics[width=0.4\textwidth]{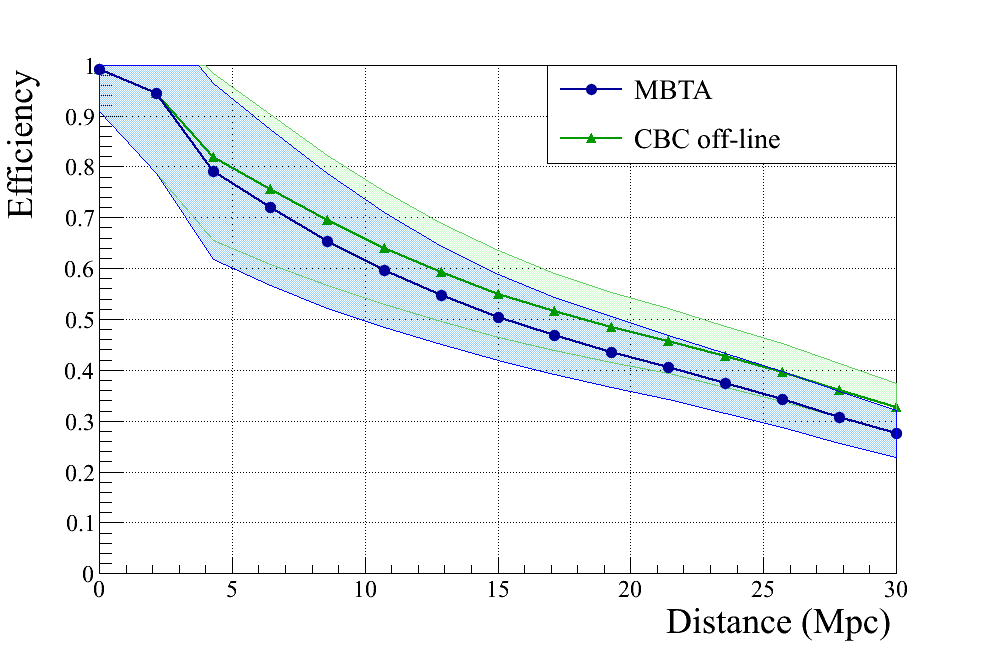}
  \caption{Detection efficiency of MBTA
    and CBC offline analysis pipeline as a function of
    distance.
Shaded regions indicate uncertainties. MBTA detects all
    simulated gravitational-wave signals injected in the nearby
    universe that the offline analysis does, but finds
    systematically less number of signals than the offline pipeline
    beyond $D > 3$\,Mpc.} \label{fig:comp2} 
\end{figure}

\paragraph{Efficiency as a function of distance}
is another key characteristic of an analysis pipeline. For each pipeline we 
measure its efficiency at recovering a GW signal as a triply coincident trigger, 
requiring also a false alarm rate less than or equal to $0.25$ events per day, 
as for the alert generation. 
The result is shown in Fig.~\ref{fig:comp2}.  At distances
of about $3$\,Mpc, MBTA begins to systematically recover fewer signals than
iHope.  As expected this trend continues as both pipelines lose efficiency at
large distances.  Much of the decrease in efficiency of MBTA can be attributed
to the fact that it imposes an effective threshold at a slightly higher SNR for a signal to be
detected. 

\begin{figure}[!h]
  \centering
  \includegraphics[width=0.4\textwidth]{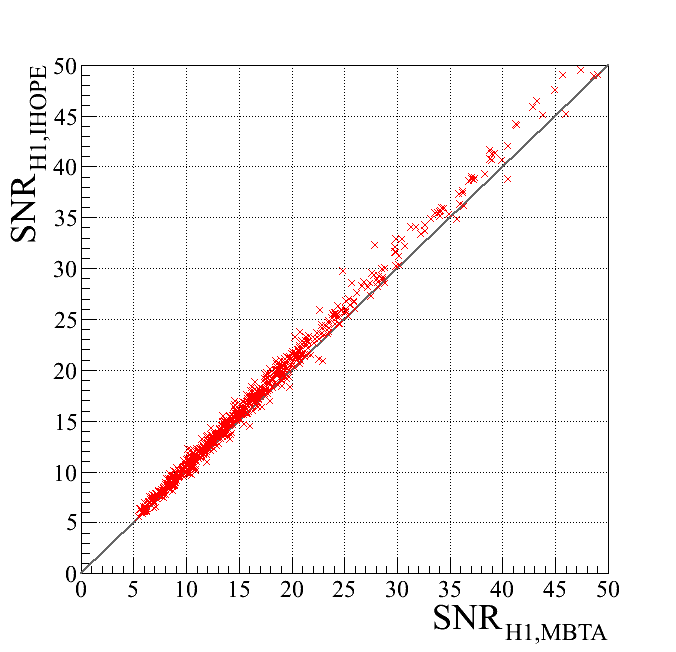}  \hspace{0.5cm}
  \includegraphics[width=0.4\textwidth]{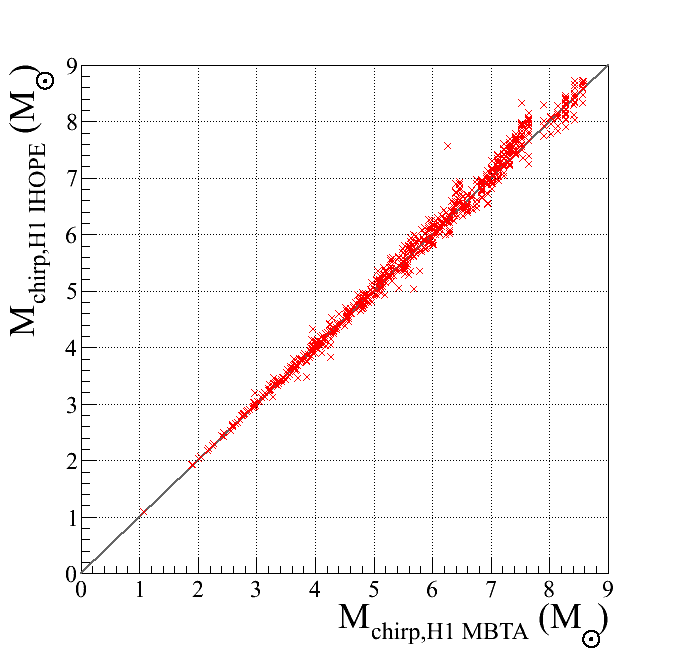}  
  \caption{Comparison of chirp mass and SNR between MBTA
    and CBC offline analysis pipeline.} \label{fig:comp3}
\end{figure}

\paragraph{Chirp mass and SNR.}
Finally, Fig.~\ref{fig:comp3} shows the SNRs and chirp masses in H1 recovered by the 
pipelines. They are in very good agreement with each other. The sparse density 
of the template bank at high mass shows up as some
discreteness in the values of chirp mass recovered by MBTA, because MBTA uses a constant
template bank, whereas iHope recomputes the noise power spectrum every $2048$\,s
and a new template bank is produced in response.
Similar agreement is found for the other two detectors.

In summary, the performance of MBTA is comparable to that of the offline
pipeline.

\section{Results}
In this section we present the results of the analysis.  We begin with an
overview of the joint LIGO-Virgo science run before turning our attention to the
triggers that passed the selection cuts.

\subsection{The S6/VSR3 run}
The joint LIGO/Virgo data taking started on August 11, 2010 when Virgo 
started its 3rd science run (VSR3), joining LIGO's 6th science run (S6). After 
a test and adjustment period during the first part of the S6/VSR3 run, the 
software of the low latency pipeline was frozen on August 27. From this 
time on, until the end of the run on October 19, the full pipeline was 
operating in production mode.
The trigger production was first monitored to validate the pipeline during 
online operations and then the submission of EM alert was enabled on September
19.

During the production period of 52.6 days, the science mode duty cycle of the 
H1, L1 and V1 detectors were respectively 63.9\%, 64.8\% and 69.7\%.   The 
three detectors were operating simultaneously in science mode for a total time
of $18.2$\,days, corresponding to a 34.6\% duty cycle. 
\par

The duty cycle of MBTA during the triple coincidence time was
94.2\%, 98.7\% and 97.8\% for the single H1, L1 and V1 triggers and 91.2\% 
generating triggers in triple coincidence.  Most of the down time occurred in
one of a small number of periods lasting a few hours.  The main problem was
temporary network overloads between the LIGO Hanford site and Virgo Cascina
site.  The resulting delay in the arrival of H1 data prevented it from being used
by MBTA.

Over the S6/VSR3 production period, 89 triple coincidences~-~including hardware 
injections~-~were detected by 
the MBTA pipeline and submitted to the GraCEDb database. A few of them (10) 
triggered multiple submissions of the same loud event corresponding to nearby
"satellite" events. Three GraCEDb submissions failed. One because of a GraCEDb
disk access problem.  The other two failures were due to a problem with
network authentication.

Future operations with improved configurations, software versions and monitoring tools 
are expected to reduce the down time of the pipeline which involved five different 
computing facilities located in Europe and North America.

\subsection{Triggers}

After removing the hardware injections, 
a total of 42 triple coincident triggers were observed during the search.
The application of the online data quality flags reduced the number of triple coincident triggers from 42 to
37.  The time coincidence window was chosen conservatively (larger than the light
travel time between sites) and, as a result, only 23 of these triggers were localizable on the sky.

At this stage of the pipeline, the triggers were passed to LUMIN 
which generated alerts for events having a false alarm rate of less than 
0.25 events per day (1 event per day up to August 31). 
This cut reduced the number of triggers to 13 which were passed 
to the control rooms and on-call experts for further quality assessments. 

Out of these 13 possible alerts generated by LUMIN, only 3 met the requirement
of having at least one neutron star ($M < 3.5\ M_{\odot}$) associated with the
merger. 
Table~\ref{tab:MBTA_triggers}  gives a snapshot of the parameters of these 
three triggers.

The first trigger, G16901, from August 31, occurred during an initial
testing period before alerts were sent out.  A decision on this trigger was
reached in the control room 14 minutes after the trigger time.
The second trigger, G20190, on September 19,
was accepted and images of the corresponding sky location 
were taken by Quest, ROTSE, SkyMapper, TAROT and Zadko.  The decision to issue
an alert was reached 39 minutes after the event occurred.
The image analysis is in progress. 
Figure~\ref{fig:skymap} shows the skymap produced for this trigger.  The $90\%$
confidence region was reduced from nearly 600 square degrees to 3.3 square
degrees with the application of the galaxy catalog.
The third trigger, G23201, on October 6, 
was unfortunately located too close to the sun, making it impossible to image.
The decision to send the trigger out occurred 16 minutes after the trigger.
Overall, these three triggers have SNR values close to the threshold value, 
with a false alarm rate of one per few days, typical of the expected background
triggers.  They are therefore not detection candidates on the basis of the 
gravitational-wave data analysis.

\begin{table*}
  \caption{\label{tab:MBTA_triggers} Parameters of the three triggers 
which passed all the selection cuts. See text for details.} 

 \begin{tabular}{cccccc}    
\hline
\hline
Detector & SNR & $D_{\rm eff}[Mpc]$ & $m_1 [M_\odot]$ & $m_2 [M_\odot]$ & ${\cal M}
[M_\odot]$  \\
\hline
  \multicolumn{6}{c}{G16901: 967254112;\, Combined SNR=9.99;\,
FAR${}^{-1}$=1.1\,days }\\
 \hline
   H1 & 6.15 & 55 & 1.03 & 2.06 & 1.26 \\
   L1 & 5.61 & 54 & 1.36 & 1.38 & 1.19 \\
   V1 & 5.52 & 19 & 1.35 & 1.37 & 1.18 \\
\hline
 \multicolumn{6}{c}{ G20190: 968932960;\, Combined SNR=10.0;\,
FAR${}^{-1}$=6.4\,days }\\
\hline
   H1 & 6.07 &  99 &  2.94 & 3.00 & 2.59  \\
   L1 & 5.65 & 106 &  3.05 & 3.11 & 2.68  \\
   V1 & 5.60 &  27 &  2.23 & 4.15 & 2.62  \\
\hline
 \multicolumn{6}{c}{ G23201: 970399241;\, Combined SNR=10.1;\,
FAR${}^{-1}$=5.5\,days }\\
\hline
  H1 & 5.75 & 58 & 1.04 & 1.99 & 1.24 \\
  L1 & 5.84 & 41 & 0.98 & 1.95 & 1.19 \\
  V1 & 5.96 & 14 & 0.97 & 1.91 & 1.17 \\
 \end{tabular} 
\end{table*}

\begin{figure}[!h]
  \centering
\includegraphics[width=8cm]{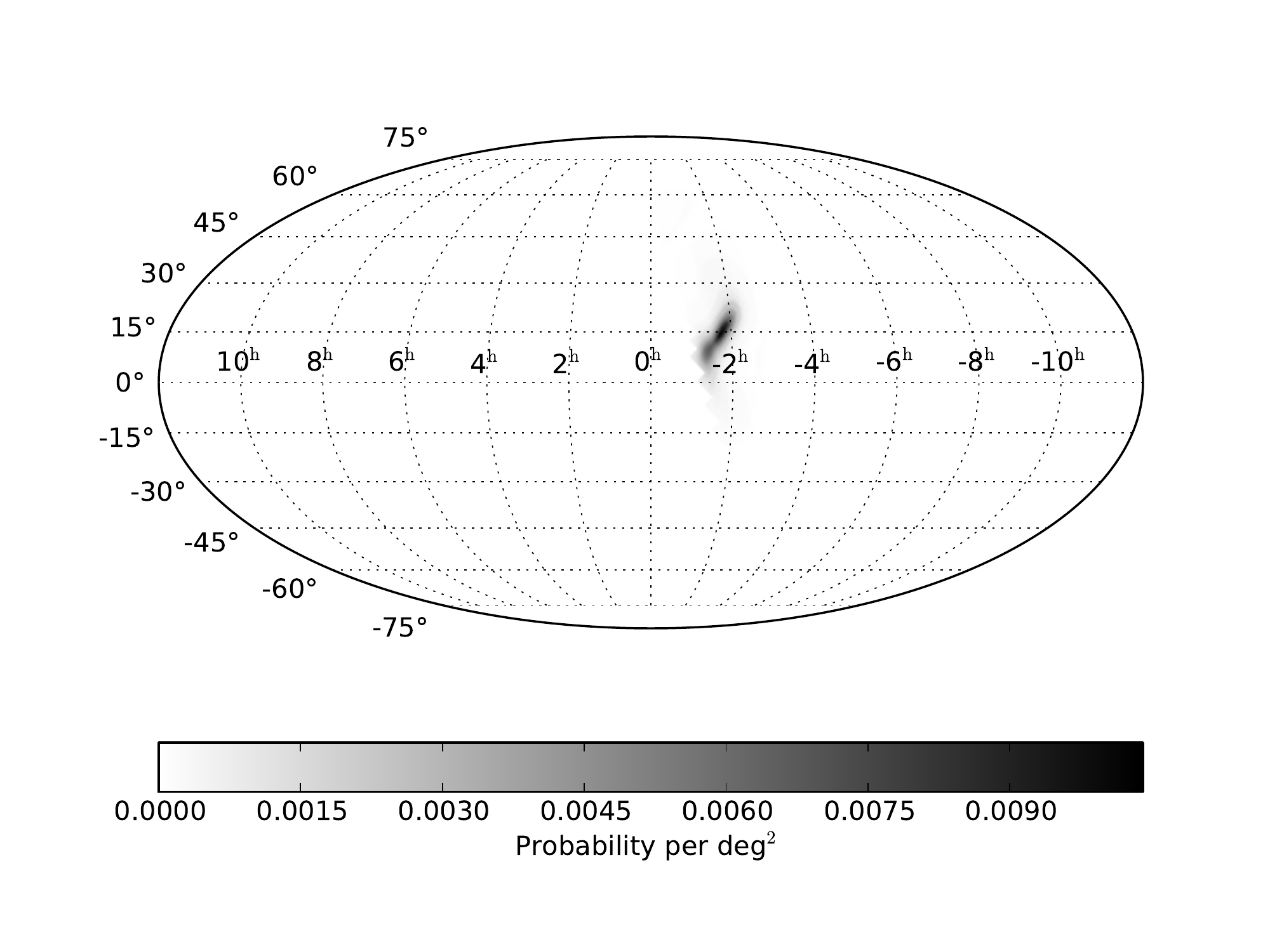}
\caption{\label{fig:skymap} Skymap for the G20190 trigger on September 19.}
\end{figure}

Figure~\ref{fig:FAR_histo} shows the cumulative rate of observed triggers as a function 
of the upper threshold applied on the estimated false alarm rate. 
The distribution focuses on triggers collected during the production
period with a FAR less than 200 per year, requiring at least one neutron star
($M < 3.5\, M_{\odot}$)
and excluding hardware injections.
The vertical line indicates the threshold of FAR lower than $\sim$ 91 per year
(0.25 per day),
which was used to determine which triggers were candidates for EM follow-up. 
This figure shows that the online FAR estimation is reasonable 
and therefore the background is under control. 
In particular, this figure shows no evidence of the FAR being underestimated, 
which is important since we do not want to unduly promote uninteresting triggers.

\begin{figure}[!h]
  \centering
\includegraphics[width=8cm]{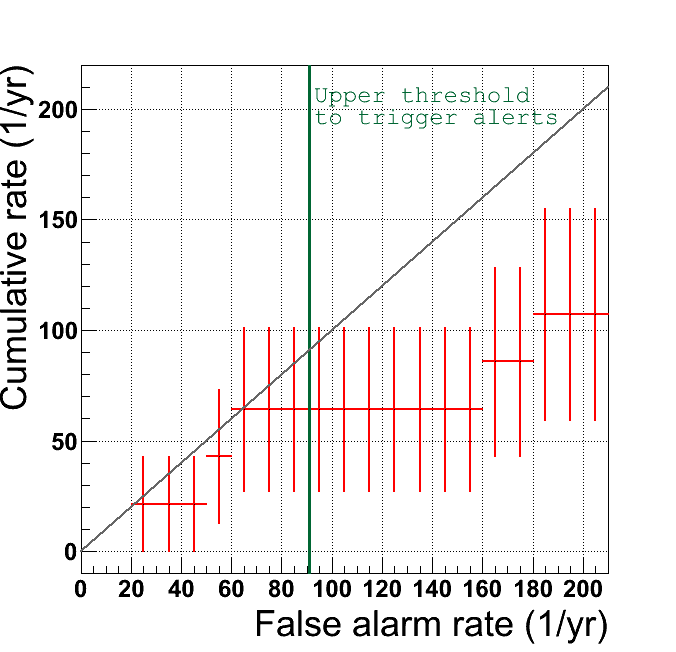}
\caption{\label{fig:FAR_histo} Cumulative rate of triggers observed during the production period (excluding 
hardware injections) as a function of the upper threshold applied on the estimated false alarm rate.}
\end{figure}

\section{Discussion}
The coalescence of binary systems containing neutron stars is the most promising
source for the detection of both gravitational and electromagnetic radiation.
We have presented the first low-latency search for the gravitational-waves from
these systems during the period between September 19 and October 20, 2010.  The
search resulted in a single trigger with a false alarm rate of one per 6.4\,days
being followed up with optical telescopes.  The results of the image
analysis are pending.

This exercise has resulted in a low-latency search for binary inspirals that
performs on par with the standard offline pipeline.  Triggers are produced on
the scale of minutes and it is likely this can be further reduced.  The limiting
latency in sending triggers to electromagnetic observatories is the human
monitoring involved.  However, our demonstration that reliable false alarm rates
can be computed rapidly suggests that this step could be removed in the advanced
detector era.

Improvements of the pipeline will be explored in the future.  For instance, the volume probed could be
optimized by applying thresholds which could depend on the sensitivity and on the data quality
of each detector. More detailed data quality information could help for this step. The search area in the sky
might be reduced by using some coherent technique, possibly in a hierarchical way. Extending the emission
of fast alerts to significant double coincidences may also be useful.

Advanced LIGO and Virgo are likely to detect $\sim$50 neutron star coalescences
per year~\citep{ratesdoc}.
Successful observation of joint EM+GW emission depends
crucially on using all the available information on these sources.  Expectations
for electromagnetic emission and their lightcurves, for example, will be
important for designing optimal observing campaigns.  In addition, even with a
three-detector network, the sky localization ability is limited.  More
detectors, such as the LCGT or some other future detector will be important for
increasing the chances of making successful joint EM+GW observations.

\begin{acknowledgements}
The authors gratefully acknowledge the support of the United States
National Science Foundation for the construction and operation of the
LIGO Laboratory, the Science and Technology Facilities Council of the
United Kingdom, the Max-Planck-Society, and the State of
Niedersachsen/Germany for support of the construction and operation of
the GEO600 detector, and the Italian Istituto Nazionale di Fisica
Nucleare and the French Centre National de la Recherche Scientifique
for the construction and operation of the Virgo detector. The authors
also gratefully acknowledge the support of gravitational-wave research by these
agencies and by the Australian Research Council, 
the International Science Linkages program of the Commonwealth of Australia,
the Council of Scientific and Industrial Research of India, 
the Istituto Nazionale di Fisica Nucleare of Italy, 
the Spanish Ministerio de Educaci\'on y Ciencia, 
the Conselleria d'Economia Hisenda i Innovaci\'o of the
Govern de les Illes Balears, the Foundation for Fundamental Research
on Matter supported by the Netherlands Organisation for Scientific Research, 
the Polish Ministry of Science and Higher Education, the FOCUS
Programme of Foundation for Polish Science,
the Royal Society, the Scottish Funding Council, the
Scottish Universities Physics Alliance, The National Aeronautics and
Space Administration, the Carnegie Trust, the Leverhulme Trust, the
David and Lucile Packard Foundation, the Research Corporation, and
the Alfred P. Sloan Foundation.
\end{acknowledgements}

\bibliographystyle{aa}
\bibliography{../bibtex/iulpapers}
\end{twocolumn}
\end{document}